\documentclass[10pt]{article}
\usepackage{wrapfig}
\usepackage{color}
\usepackage{graphicx}
\usepackage[colorlinks=true, pdfstartview=FitV, linkcolor=blue, citecolor=blue,urlcolor=blue]{hyperref}

\usepackage{amsbsy}
\usepackage{amssymb}
\def\D{\mathbb D}

\usepackage{verbatim}

\renewcommand{\theequation}{\arabic{section}.\arabic{equation}}

\makeatletter
\@addtoreset{equation}{section}
\makeatother

\newtheorem{theorem}{Theorem}[section]

\newtheorem{exercise}{Exercise}[section]

\newtheorem{lemma}{Lemma}[section]
\newtheorem{remark}{Remark}[section]

\newtheorem{proposition}{Proposition}[section] 
\newtheorem{corollary}{Corollary}[section] 
\newtheorem{definition}{Definition}[section]
\def\le{\left}
\def\ri{\right}
\def\ds{\displaystyle}

\def\res{\mathop{\mathrm {res}}\limits_}

\def\br{\begin{remark}}
\def\er{\end{remark}}
\def\bt{\begin{theorem}}
\def\et{\end{theorem}}
\def\bc{\begin{corollary}}
\def\ec{\end{corollary}}
\def\bx{\begin{examp}\small}
\def\ex{\end{examp}}
\def\bxr{\begin{exercise}\small}
\def\exr{\end{exercise}}
\def\bl{\begin{lemma}}
\def\el{\end{lemma}}
\def\bd{\begin{definition}}
\def\ed{\end{definition}}
\def\bp{\begin{proposition}}
\def\ep{\end{proposition}}
\def\be{\begin{equation}}
\def\ee{\end{equation}}
\def\ov {\overline}

\def\&{\hspace{-15pt}&}
\def\bea{\begin{eqnarray}}
\def\eea{\end{eqnarray}}
\def\beas{\begin{eqnarray*}}
\def\eeas{\end{eqnarray*}}
\def\B{\mathbf B}
\def\A{\mathbf A}
\def \pa{\partial}
\def\C{{\mathbb C}}
\def\L{\mathcal L}
\def\R{{\mathbb R}}
\def\N{{\mathbb N}}

\def\wh{\widehat}

\def\Z{{\mathbb Z}}

\def\a{\alpha}
\def\d{\,\mathrm d}

\def\1{{\bf 1}}
\def\wt{\widetilde}

\date{}
\begin{document}
\baselineskip 15pt plus 1pt minus 1pt

\vspace{0.2cm}
\begin{center}
\begin{Large}
\fontfamily{cmss}
\fontsize{17pt}{27pt}
\selectfont
\textbf{
First Colonization of a Spectral Outpost in Random Matrix Theory
}
\end{Large}\\
\bigskip
\begin{large} {M.
Bertola}$^{\ddagger,\sharp}$\footnote{Work supported in part by the Natural
    Sciences and Engineering Research Council of Canada
(NSERC).}\footnote{bertola@crm.umontreal.ca}, S. Y. Lee$^{\sharp}$.
\end{large}
\\
\bigskip
\begin{small}
$^{\ddagger}$ {\em Department of Mathematics and
Statistics, Concordia University\\ 1455 de Maisonneuve W., Montr\'eal, Qu\'ebec,
Canada H3G 1M8} \\
$^{\sharp}$ {\em Centre de recherches math\'ematiques\\ Universit\'e\ de
Montr\'eal } \\
\end{small}
\bigskip
{\bf Abstract}
\end{center}
We describe the distribution of the first {\em finite} number
of eigenvalues in a newly-forming band of the spectrum of the random
Hermitean matrix
model.
The method is rigorously based on the Riemann--Hilbert analysis of the
corresponding orthogonal polynomials. 
We provide an analysis with an error term of order $N^{-2\gamma}$ where $1/\gamma = {2\nu+2}$ is the exponent of non-regularity of the effective potential, thus improving even in the usual case the analysis of the pertinent literature.

The behavior of the first finite number of zeroes (eigenvalues) appearing in the new band is analyzed and connected with the location of the zeroes of certain Freud polynomials. In general all these newborn zeroes approach the point of nonregularity at the rate $N^{-\gamma}$ whereas one (a {\em stray zero}) lags behind at a slower rate of approach.
The kernels for the correlator functions in the scaling coordinate near the emerging band are provided together with the subleading term: in particular the transition between $K$ and $K+1$ eigenvalues is analyzed in detail. 
\vspace{0.7cm}

{Keywords: \parbox[t]{0.8\textwidth}{Orthogonal polynomials, Random matrix theory, Schlesinger transformations,\ Riemann--Hilbert problems.}}
\vskip 15pt
{AMS-MSC2000: 05E35, 15A52}

\tableofcontents

\section{Introduction} 
In this paper we consider the Hermitean matrix model in the scaling regime or
--which is the same-- the orthogonal polynomials on the real line with a varying
weight, in the same spirit as \cite{BleherIts, DKMVZ}. We address a particular situation of ``nonregular'' or ``critical'' potential: this means that the mean--field electrostatic potential vanishes at some point $\xi_0$ outside of the support of the equilibrium measure. This situation corresponds to a recent investigation \cite{EynardBirth} and is the situation where a band in the spectrum of the corresponding matrix model or a new component of accumulation of the zeroes of the orthogonal polynomials is about to appear (or has just disappeared).  

In the paper we will be mostly taking the point of view of approximation theory and hence we will focus on the orthogonal polynomial side, but the more physically--oriented reader will have no difficulty in translating those results; in a picturesque way we will think of the zeroes of the polynomials as a growing  population and thus call ``colonies'' the first zeroes appearing near the new band, also termed the ``outpost''. This should explain the catchy title.

We will be using a particular (simplified) version of the double--scaling limit: in this approach we keep the potential $V(x)$ and the total charge $T$ {\em fixed} but we add a piecewise constant perturbation of order   $ \ln (N)/N$ to the potential near the outpost. In due time (App.  \ref{App1}) we will explain how this simplified approach yields in fact identical results to the usual double--scaling.

In order to explain in more detail the setup, suppose that the effective potential \cite{SaffTotik} $\varphi(x) = \frac 1T (V(x) - 2g + \ell)$ vanish at $\xi_0$ as $C(x-\xi_0)^{2\nu+2}$ (with some $C>0$); here $\xi_0$ is some point outside of the support of the 
equilibrium measure \cite{SaffTotik}.  We then modify the potential (Sect. \ref{SectChemical})  by adding a step-like perturbation of the form
\be
V(x)\to \wt V(x) = V(x) -\frac {2\varkappa T \gamma}{N}\ln N \chi_J(x)\ ,\qquad \gamma:=\frac 1{2\nu+2}\ .
\ee
Here $\chi_J$ is the characteristic function of a small interval around $\xi_0$.
The real parameter $\varkappa$ determines the strength of the perturbation and the constants are crafted for later convenience.

For $\varkappa<0$ the orthogonal polynomials do not exhibit any peculiar behavior in the large $N$ limit; for positive values of  $\varkappa$ new zeroes of the OPs start appearing near the outpost. It is natural that --since we can have only an integer number of such roots-- there are {\em transition points} in the asymptotic behavior for special values of $\varkappa$. Specifically the normalizations we have chosen are such that 
in the asymptotic regime there are $K$ roots near the outpost, where $K$ is the integer nearest to $\varkappa$. Clearly transitions must occur at $\varkappa\in \N + \frac 12$. 
The phenomenon is already captured by the {\em leading order asymptotics} (Sect. \ref{SectChemical}): indeed one can construct a uniform  approximation  to the orthogonal polynomials to order $N^{-\gamma(1-2|\delta|)}$. The  approximation is best when $\varkappa$ is an integer and gets progressively worse and worse as $\varkappa$ approaches a half--integer. 
In particular when $\varkappa$ is a half integer the approximation breaks down (the error term is no longer vanishing as $N\to\infty$). This apparent obstruction was noted in \cite{EynardBirth} (see also \cite{Claeys, MoBirth}).

We then show how to obtain an improved approximation in Sect. \ref{sectimproved}; indeed we construct a uniform asymptotic solution with an error term of order $N^{-2\gamma}$ (uniformly in $\varkappa$!). Even in the ordinary case $\varkappa=0$ (which was dealt with in \cite{DKMVZ}) our approximation is better than the one usually provided in the literature (in fact in  Sect \ref{arbitrary} we show how to obtain an approximation of order $N^{-1}$ for arbitrary $\varkappa$).

Using this information we can study in detail the {\bf asymptotic behavior of the first roots} (Sect. \ref{sectroots}); in a scaling parameter $\zeta \sim N^\gamma (z-\xi_0)$ they are related to the roots of the {\em Freud's polynomials} for the weight ${\rm e}^{-\zeta^{2\nu+2}}\d \zeta$; in particular 
\begin{itemize}
\item if $\varkappa \in [K, K+1/2)$ then  they are within a distance $\mathcal O(N^{-2\gamma \delta})$ from the roots (in the $\zeta$--coordinate) of the $K$-th Freud polynomial.  
\item if $\varkappa \in (K-1/2,K)$ then there are still $K$ roots; however while $K-1$ of them are within $\mathcal O(N^{-2\gamma\delta})$ from the roots of the $(K-1)$-st Freud polynomial, the ``last one'' lags behind and meanders at a distance $\mathcal O(N^{2\gamma\delta})$ in the $\zeta$--coordinate. Note that --while escaping to infinity in the $\zeta$--scaling parameter, such root is actually converging to the outpost at a rate $N^{-\gamma+2\gamma \delta}$.
\item when $\varkappa \in \N+1/2 $ (namely $\delta = \frac 12$) then there is a stray root that remains at a finite distance from the outpost, while the remaining converge to it. 
\end{itemize}

In the main body of the paper we make the simplifying assumption that the support of the equilibrium measure consists of one interval ({\bf one--cut assumption}). However this is only a simplification and none (or almost) of the conclusions are at all dependent on it, but of course the formul\ae\ for the outer parametrix are much simpler to write and easier to handle also for those readers who do not know well the theory of Theta functions.

In App. \ref{multicut} we show (in a somewhat sketchy form) how to generalize to an arbitrary number of cuts: only one detail cannot be fully addressed in this general case, and concerns with the improved asymptotic for some of the exceptional values of $\varkappa\in \N+1/2$ and exceptional spectral curves, namely the equivalent of formula \ref{detB} for the multi--cut case and the corresponding sign. The knowledge of the sign of (the imaginary part of)  (\ref{detB}) is necessary to ensure that in particularly exceptional circumstances certain denominators (\ref{FG}) do not vanish.

\br
In a strange twist of events while the present manuscript was in the latest phases of preparation, a similar preprint \cite{Claeys} has appeared where the author uses a RH analysis for the simplest nonregular case. Shortly (two days) after another independent preprint \cite{MoBirth} on the same topic has appeared, dealing with a more general type of nonregularity (of the same type we deal). 

Our work however provides a refined error analysis up to order $N^{-2\gamma}$ whereas both the previous papers apparently give only the leading term asymptotics, with an error term of order $N^{-\gamma(1+2\delta)}$, thus not valid at the transition points (although in \cite{Claeys} an analysis of the  half--integer case is also provided but only for the simplest nonregularity exponent).

In addition we provide a detailed analysis of the location of the zeroes of the orthogonal polynomials near the outpost.
\er
\section{General setting}
The setting of the present paper will be identical in the most part to
\cite{DKMVZ}, \cite{BleherIts}.  Although we present in a
self-contained way we refer the reader to the pertinent literature for
more details on the subject.

Consider the Hermitean matrix model with measure given by
\be
\label{matrixmeasure}
\frac 1{Z_N} {\rm e}^{-\frac NT {\rm tr} V(M)} {\rm d} M
\ee
where $Z_N$ is a normalization constant.
It is known \cite{MehtaBook} that the model is ``solvable'' in terms of orthogonal polynomials (OPs) and that all spectral statistics can be described in terms of suitable kernels constructed in terms of OPs.

Let $\{p_n(x)\}$ be the corresponding (monic) OPs that satisfy the following orthogonality condition with the potential $V(x)$ which we assume to be {\bf real} and {\bf analytic}.
\be\label{orthogonality}
\int_{\R} p_n(x) p_m(x) {\rm e}^{-\frac NT V(x)}\d x = h_n\delta_{nm}.
\ee
The spectral statistics of the model is determined by the Christoffel--Darboux kernel \cite{MehtaBook}
\be
K(x,x') = \sum_{j=0}^{n-1} \frac {p_j(x)p_j(x')}{h_j} =  \frac {
p_n(x)p_{n-1}(x')- p_{n-1}(x)p_{n}(x')}{h_{n-1}(x-x')}
\ee

The OP are uniquely characterized by the following Riemann--Hilbert problem.
Define for $z\in \C \setminus \R$ the matrix
\be
Y(z):= Y_n(z):= \le[
\begin{array}{cc}
p_n(z) & \phi_n(z)\\
\frac {-2i\pi}{h_{n-1}}  p_{n-1}(z) & \frac {-2i\pi}{h_{n-1}}\phi_{n-1}(z)
\end{array}
\ri]\ ,\qquad \phi_n(z):= \frac 1 {2i\pi} \int_\R \frac {p_n(x) {\rm e}^{-\frac
NT V(x)}\d x}{x-z}.
\label{OPRHP1}
\ee
The above matrix has the following jump-relations and asymptotic behavior that
{\bf uniquely characterize it} \cite{FIK0, FIK1, FIK2, FIK3} (we drop the explicit dependence on
$n$ for brevity)
\bea
Y_+(x) = Y_{-}(x) \le[
\begin{array}{cc}
1 & {\rm e}^{-\frac NTV(x)}\cr
0&1
\end{array}
\ri]\ ,\qquad
Y(z) \sim\big(\1 + \mathcal O(z^{-1})\big) \le[\begin{array}{cc}
z^n &0\cr 0&z^{-n}
\end{array}\ri].
\label{OPRHP2}
\eea
Replacing the orthogonality condition (\ref{orthogonality}) by the above jump (and boundary) conditions (\ref{OPRHP2}) we obtain the Riemann--Hilbert problem for the OPs. 
Using this setup we especially want to investigate the asymptotics of the OPs as their degree $n:=N+r$ goes to infinity while $r$ being fixed to an integer.

The first step to solve the problem is to find the {\bf $g$-function}, which we will define using the {\bf equilibrium measure} below.
We briefly recall that, in the simple case where the contour of integration in (\ref{orthogonality}) is the real axis (see \cite{BertolaMo, BertoBoutroux} for a general approach not relying on a variational problem), the equilibrium measure is obtained from the solution of a variational problem for a functional over probability measures on the real axis, in the sense of potential theory \cite{SaffTotik}.
Indeed define the {\bf weighted electrostatic energy} \cite{SaffTotik}
\be
\mathcal F[\mu] := 2\int_\R V(x) \d \mu(x)  +  \int_\R \int_\R\ln \frac
1{|x-x'|} \d\mu(x)\d\mu(x') 
\ee
where $\d\mu$ is a positive measure supported on the real axis with
total mass $T= \int_\R \d \mu(x)$.

It is known that the functional $\mathcal F$ attains a unique minimum (under
mild assumptions on the growth of $V(x)$ at infinity) at a measure $\rho$ that
is called the {\bf equilibrium measure} {\cite{SaffTotik,Deift}}.

 It is also known \cite{McLaughlinDeiftKriecherbauer} that  the
support of the measure $\rho$ consists of a finite union of  disjoint bounded
intervals and that $\rho$ is smooth on the interior of the support.

Taking avail of the equilibrium measure, the $g$ function \cite{Deift} is then typically defined as
\be
g(z):= \int_\R \rho(x) \ln (z-x) \d x=T\ln z +{\cal O}(z^{-1})\label{gfunct},
\ee
where the logarithm must be defined with an appropriate cut extending --say-- from the leftmost endpoint of the support of $\rho$ to $+\infty$. The derivative of the function $g(z)$ (the ``resolvent'')  satisfies a pseudo--algebraic equation which is key to many considerations in a different context but will be mostly irrelevant in this paper.

The main properties that enter the steepest descent analysis are the standard
properties of the logarithmic transform. To this end we
 note that the representation (\ref{gfunct}) implies immediately that $\Re g(x)$ is
harmonic away from the support of $\rho$ and  continuous on the whole complex
plane. {The Euler--Lagrange variational equations equivalent to the optimality of the equilibrium measure $\rho$ \cite{SaffTotik} can be rephrased in terms of the following conditions for the $g$--function}
\begin{itemize}
\item for $x\in \R$ we have
\be
\Re \varphi(x)\geq 0 \label{ineq}, \quad\varphi(z):=\frac{V(z)}{2}-g(z)+\frac{\ell}{2} = \frac{V(z)}{2}-\int \rho(y) \ln (x-y) \d y +\frac{\ell}{2}
\ee
for a suitable real constant $\ell$.  $\Re\varphi$ is the effective potential of the related electrostatic problem.
\item The opposite inequality (and hence the equality) holds on the support of $\rho$.  Especially, $\ell$ is chosen such that $\Re\varphi=0$ on the support of $\rho$. (The support of $\rho$ will be called the {\bf cuts} because they form the cuts of the functions $g'(z)$ and $\varphi'(z)$.) {Here and in the previous point, the $g$--function should be understood as the analytic function  defined by its integral representation (\ref{gfunct}) on the simply connected domain obtained by removing a half--line starting e.g. at the rightmost endpoint of the support of $\rho$ and extending towards $-\infty$. Then $\Re \varphi(x)$ is actually nothing but the boundary-value $1/2(\varphi_+(x) + \varphi_-(x)) $}
\item In suitable finite left/right neighborhoods of the cuts, the function $\Re \varphi(x)$, which is
also harmonic in the domain of analyticity of $V(x)$, is {\bf negative}.
\end{itemize}

The situation we want to address in this paper is the case where the inequality (\ref{ineq}) is not strict on $\R$ outside the cuts, but at some point $\xi_0$ {\bf outside} of the cuts, the inequality is an equality.   This situation corresponds to the critical situation where a cut is about to emerge (or has just disappeared) at $\xi_0$. In an optimistic view, we will look at the situation as of that of an {\em emerging} spectral band, being gradually populated by eigenvalues; it is thus appropriate to refer to the neighborhood of the emerging band as an {\bf outpost colony} of eigenvalues.

The pair of potentials $V(x)$ and total charge $T$ for which the inequality (\ref{ineq}) is not strict either outside or inside the cuts (this last occurrence corresponding to the {\em merger} of two cuts) are called {\bf nonregular} or, in the more physical oriented literature, {\bf critical}. The steepest descent analysis was completely carried out in \cite{DKMVZ} in general terms and specific analysis linking with Painlev\'e\ theory was carried out in \cite{BleherIts} in the case of  a merger. For the ``birth of a cut'' (i.e. outpost colonization), a heuristic arguments and a double--scaling approach\footnote{This means that not only $n \to \infty$ but also $V,T$ are let
depend on $n$ in a ``slow'' and fine--tuned way.}
were used  in \cite{EynardBirth}; while the intents of our note and of \cite{EynardBirth} are clearly the same, the methods employed are radically different.

The conclusion that we achieve in rigorous mathematical way will be --however-- parallel to that of \cite{EynardBirth}, namely to show that we can describe the statistics of the first {\em finite} number of eigenvalues  that populate the forming cut in terms of an effective ``microscopic'' matrix model of the size of the population of the outpost. In addition we strengthen those result by localizing exactly the relevant, finite number of roots of the orthogonal polynomials.
\section{Modified setting: changing chemical potential}
\label{SectChemical}
\begin{wrapfigure}{r}{0.5\textwidth}
\resizebox{0.45\textwidth}{!}{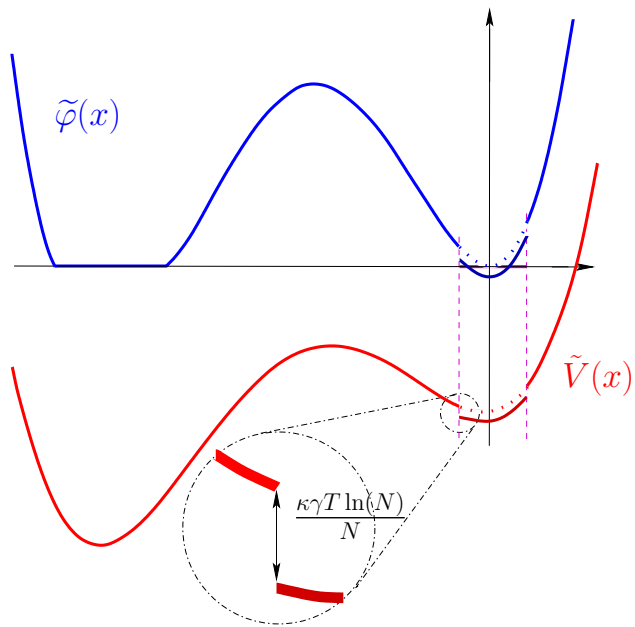}
\caption{The potential $\wt V$ with the chemical potential added, and the corresponding effective potential.}
\label{figone}
\end{wrapfigure}
In order to study a nontrivial scaling limit we modify the setting of the problem as follows.
As mentioned in the previous section, we consider the situation when we have $\Re\varphi(\xi_0)=0$ for $\xi_0\in {\mathbb R}$ outside the cuts.
Suppose that $\xi_0$ is the point where $\varphi(z)$ has the critical point of order $2\nu+2$, i.e. 
\be
\varphi(z) =  \mathcal O\Big((z-\xi_0)^{2\nu+2}\Big),\ \ z\sim \xi_0.
\ee

Then we choose a finite open interval $J$ containing $\xi_0$ that does not contain any other turning points.  We then consider the following {\bf modified orthogonality relation}:
\bea\label{modOR}
h_{nm} \delta_{nm} =
\int_{\R\setminus J} p_n(x)p_m(x) {\rm e}^{-\frac NT V(x)}\d x  +\cr
+N^{2\varkappa \gamma} \int_{ J} p_n(x)p_m(x) {\rm e}^{-\frac NT V(x)}\d x,\cr
\eea
where we have defined the {\bf exponent of nonregularity} 
\be\gamma:= (2\nu + 2)^{-1}.\ee
In the above relation the parameter  $\varkappa\in{\mathbb R}$ will eventually determine the size of the population of the colony near $\xi_0$. 

As the reader may realize this amounts simply to a step-wise modification of the potential: if $\chi_{J}$ is the characteristic function of the interval $J$ then we may rewrite (\ref{modOR}) as a single integral without $N^{2\varkappa\gamma}$ using the modified potential
\be
\wt V(x) := V(x) - \le(\frac {2\varkappa T\gamma }{N} \ln N\ri) \chi_J(x).
\ee
The most interesting regime will turn out to be $\varkappa >0$, so that the potential is slightly depressed near $\xi_0$ (Fig. \ref{figone}).
While this is a ``small'' perturbation of the potential (which would be irrelevant in a noncritical situation), since our potential is critical the effect of this perturbation is fine-tuned to obtain a nontrivial perturbation. 
It will also become clear that the actual choice of $J$ is irrelevant as long as it  contains $\xi_0$ and no cuts.

Although this ``discontinuous'' deformation may seem quite artificial at first, it should become apparent later on that it actually makes no difference on the actual behavior near the outpost. 
In a certain sense this is the essence of universality, but we will explain in Appendix \ref{App1} how to approach the same problem from a  more ``canonical'' double--scaling limit, while retaining the main features. 

The advantage of this simplified approach is that allows us to immediately concentrate on the significant features (the actual RHP) without hindering the analysis into details regarding the appropriate $g$--function.


\subsection{Normalized and lens-opened RHP}

\begin{wrapfigure}{l}{0.7\textwidth}
\resizebox{0.7\textwidth}{!}{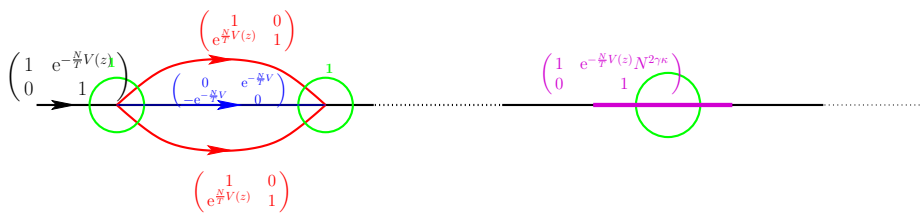}
\caption{\label{figure_Ylens}The jump matrices for $Y$.}
\end{wrapfigure}

Taking avail of the general wisdom, in order to streamline the derivation we open the lenses {\bf before} normalizing the problem\footnote{We are of course assuming that $V(z)$ is real-analytic.}, thus modifying the jumps as shown in figure \ref{figure_Ylens}.  Lens opening simply means that we {\em redefine} 
\bea
&&Y_{\mbox{new}}:=
Y\le[\begin{array}{cc}\scriptstyle1&\scriptstyle0\\\scriptstyle-{\rm e}^{\frac{N}{T}V(z)}&\scriptstyle1\end{array}\ri],\quad\mbox{ on the upper lip},\\
&&Y_{\mbox{new}}:=Y\le[\begin{array}{cc}\scriptstyle1&\scriptstyle0\\\scriptstyle{\rm e}^{\frac{N}{T}V(z)}&\scriptstyle1\end{array}\ri],\quad\mbox{ on the lower lip}.
\eea

For the time being the jumps on the green circles are the identity but later we define separate RHP problems inside the circles.  Then we will call all the RHP inside one of the green disks the {\bf local problem} whereas we call the problem outside the {\bf outer problem}.

After the lens-opening we define, 
\be\label{Ytilde}
\wt Y(z):=  {\rm e}^{\frac {N\ell} {2T}\sigma_3}Y(z) {\rm e}^{-\frac N T g(z)\sigma_3}{\rm e}^{-\frac {N\ell}{2T}\sigma_3},
\ee
which satisfies a new, simpler RHP:
\bea\label{RHPY1}
\wt Y(x)_+ &\&= \wt Y(x)_- \le[
\begin{array}{cc}
\ds {\rm e}^{\frac NT(g_--g_+)} & \ds {\rm e}^{-\frac NT(V-g_+- g_-+\ell)}N^{2\varkappa\gamma \chi_J(z)}\\
0 & \ds {\rm e}^{\frac NT(g_+-g_-)}
\end{array}
\ri]
\\\label{RHPY2}
\wt Y(z)&\& \simeq (\1 + \mathcal O(z^{-1})) z^{ r\sigma_3}\ ,\qquad z\sim \infty
\eea

\begin{wrapfigure}{l}{0.7\textwidth}
\resizebox{0.7\textwidth}{!}{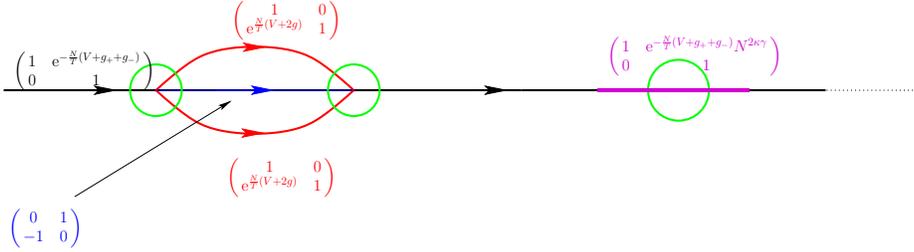}
\caption{\label{figure_Ytildelens} The jump matrices for $\tilde Y$.}
\end{wrapfigure}

For simplicity we assume that there is only one finite band in the spectrum, namely the spectral curve is of genus $0$; the generalization to more bands is not conceptually a problem but requires the use of $\Theta$--functions which would make the note quite more technical and long.
Under this assumption, for $x\in{\mathbb R}$,
\bea
g_+(x) &\& = -g_-(x)+V(x)+\ell \ ,\qquad \hbox{ for $x\in\R$ on the cut},\\
g_+(x) &\& = g_-(x) - 2i\pi T \ ,\qquad \hbox {for $x\in\R$ on the right of the cut},\\
g_+ (x)&\& = g_- (x)\ ,\qquad \hbox {for $x\in\R$ on the left of the cut}.
\eea
Everywhere else on the complex plane $g(z)$ is holomorphic.
On account of these properties for the $g$--function the jumps for $\wt Y$ are shown in the figure \ref{figure_Ytildelens}.

In the following the size of the green circles will be fixed to a nonzero value.
In this case the reader can verify that --{\em outside of the green disks}-- the jumps on the black and red lines become exponentially close to the identity, and uniformly so  in $L^2\cap L^\infty$.

\subsection{Outer parametrix}\label{section_outerparametrix}

For simplicity of exposition  we assume that there is only one spectral band ($1$-cut) apart from the one that is about to emerge. Furthermore we assume that the irregular point of the potential problem is set to $\xi_0=0$, without loss of generality.

\begin{wrapfigure}{r}{0.7\textwidth}
\resizebox{0.7\textwidth}{!}{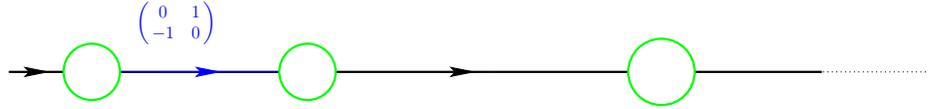}
\caption{Jump matrix for $\Psi$.\label{figure_Spinor}}
\end{wrapfigure}
Removing all the jumps that are exponentially close to the identity, we are left with the jump matrix as shown in Figure \ref{figure_Spinor}. 
This provides the asymptotic RHP that we will use to define {\bf outer parametrix}.
Below we describe the RHP that the outer parametrix $\Psi$ satisfies.
(For a specific solution to the RHP we will use $\Psi$ with a subscript or a dressing such as in $\wt\Psi_K$.)
\bea\label{outerRHP1}
&&\Psi(z) \simeq  (\1 + \mathcal O(z^{-1})) z^{r\sigma_3},\quad z\sim\infty, 
\\\label{outerRHP2}
&&\Psi(z)_+  = \Psi (z)_-\le[\begin{array}{cc}
0 & 1\\
-1 & 0
\end{array}
\ri],\quad \mbox{on the cut}.
\eea
It also needs to be supplemented by the boundary conditions at the turning points,
\be\label{outerRHP3}
\Psi(z) = \mathcal O\big( (z-a)^{-\frac 14}\big),\quad
\Psi(z) = \mathcal O\big( (z-b)^{-\frac 14}\big),\ee
where $a$ and $b$ are the two turning points.
For the specific outer parametrix $\Psi_K$ that we consider soon, the growth condition at the outpost is given by
\be\label{outerRHP4}
\Psi_K(z) = [\A_K,\B_K]\, z^{K\sigma_3}
:=\le[\begin{array}{cc}A_x(z)&B_x(z)\\A_y(z)&B_y(z)\end{array}\ri]\, z^{K\sigma_3}.
\ee
The matrix $[\A_K,\B_K]$ is analytic at $z=0$ (note that $\xi_0=0$) and $\det[\A_K,\B_K]=1$.   The above four conditions give the Riemann--Hilbert problem for $\Psi$.
We now describe a specific solution $\Psi_K$ (and $[\A_K,\B_K]$) to the above RHP.

We now use the fact that there is only one cut so that the two--sheeted cover of the $z$--plane is a rational (i.e. genus $0$) curve. The modifications needed for the case of an arbitrary number of cuts are sketched in appendix.

Let $t$ be the uniformizing map of the genus-$0$ Riemann surface.  We let $t_0$ on the $t$-plane to map to the outpost on $z$-plane.  For the simplicity of the normalization we choose the location of the cut and the outpost in the following way.
\be
z(t) := \frac{b-a}{4}\left(t + \frac 1 t\ri)+\frac{b+a}{2} = \frac{a-b}{4t_0}(t-t_0)\le(\frac 1t-t_0\ri),
\ee
where $a<b$ are the endpoints of the band; in the $t$--plane they correspond to $t=\pm 1$. There are $2$ points in the $t$--plane projecting to $z=0$ namely the two solutions of $z(t)=0$. We denote the one outside the unit circle by $t_0$, the other being $\frac 1 {t_0}$.

\begin{wrapfigure}{r}{0.5\textwidth}
\resizebox{0.48\textwidth}{!}{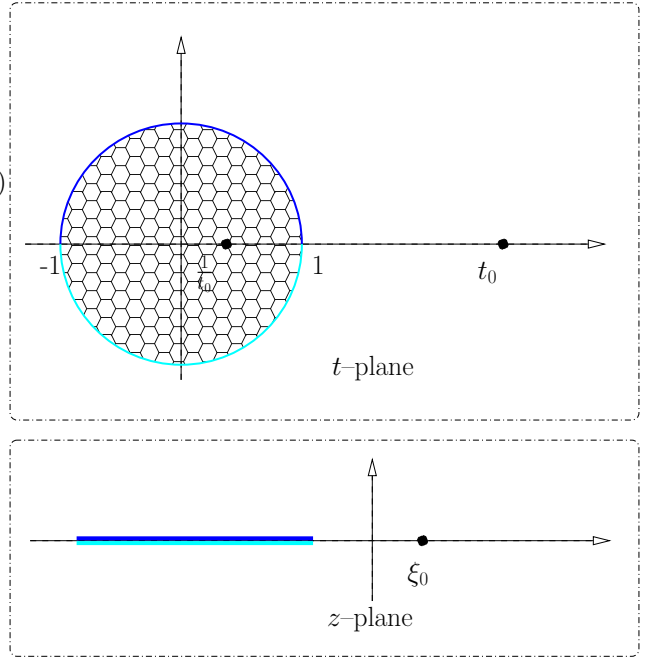}
\caption{The uniformization of the plane sliced along the support of the equilibrium measure. The pattern region is the ``unphysical sheet''.}
\end{wrapfigure}

Define the {\bf spinorial Baker--Akhiezer vectors}
\bea\label{spinor}
\Psi^{(1)} (t):= \le[
\begin{array}{c}
t^r\le(\frac{t-t_0}{t -1/t_0}\ri)^K\sqrt{\d t}\\
-i t^{r-1}\le(\frac{t-t_0}{ t- 1/t_0}\ri)^K \sqrt{\d t}
\end{array}
\ri]\ ,\cr \Psi^{(1),\star} (t):= \Psi^{(1)} \le(\frac 1 t\ri) = \le[
\begin{array}{c}
i t^{-r-1}\le(\frac{ t-1/t_0}{t- t_0}\ri)^K \sqrt{\d t}\\
 t^{-r}\le(\frac{t - 1/t_0}{t- t_0}\ri)^K \sqrt{\d t}
\end{array}
\ri]
\eea
where $K$ is the {\bf closest nonnegative integer} to $\varkappa $ (e.g. if $\varkappa =2.4$ then $K=2$, if $\varkappa = 2.6$ then $K=3$). Note that the definition is ambiguous for $\varkappa \in \frac 1 2+\Z$; indeed it will be seen that for these exceptional values we cannot obtain a  strong asymptotic result using these methods and we need to use a refinement (Sect. \ref{sectimproved}),  and the asymptotic for of the OPs has a discontinuous change, namely the model exhibits a {\bf nonlinear Stokes' phenomenon} in $\varkappa $.

The advantage of this spinor representation and the uniformizing coordinate is that we can easily write a general solution to the RHP  (\ref{outerRHP1}),(\ref{outerRHP2}),(\ref{outerRHP3}),(\ref{outerRHP4}) of the outer parametrix when there are exceptional points such as the outpost.

Using (\ref{spinor}) one can write the following solution.
\bea\label{outerK}
\Psi_K(t(z) ):=\frac { \le(\frac {(b-a)}4\ri)^{r\sigma_3+\frac 1 2} }{\sqrt{\d z}} \le[\Psi^{(1)}(t(z)) ,  \Psi^{(1),\star}(t(z))\ri]  = \frac  { \le(\frac {(b-a)}4\ri)^{r\sigma_3} }{\sqrt{\frac{4 z'(t)}{b-a} }}\le[
\begin{array}{cc}
t^r & \frac {i} {t^{r+1}}\\
-it^{r-1} & \frac 1 {t^{r}}
\end{array}
\ri]
\le(\frac{t-t_0}{t-  1/t_0 }\ri)^{K\sigma_3}\bigg|_{t=t(z)}
\eea
In the above $t(z)$ is the determination of $t$ that lies outside the unit circle for $z$ not on the cut (the {\bf physical sheet}), and the appropriate value on the cut viceversa.  The squareroot is the determination that behaves as $\sqrt{\frac{4z'(t)}{b-a}}\sim1$ near $t=\infty$.

We see also that the above matrix behaves as follows for large $x$ and has the following jump-discontinuity
\bea
\Psi_K(z) &\& \simeq  (\1 + \mathcal O(z^{-1})) z^{r\sigma_3}\\
\Psi_K(x)_+ &\& = \Psi (x)_-\le[\begin{array}{cc}
0 & 1\\
-1 & 0
\end{array}
\ri]
\eea

It is to be noted that 
\be
\Psi_K(z) = \mathcal O( (z-a)^{-\frac 14}) \ ,\qquad 
\Psi_K(z) = \mathcal O( (z-b)^{-\frac 14})\ .
\ee

More importantly, the parametrix we have constructed behaves as follows in the vicinity of the outpost
\be
\Psi_K(z) = \le(C + \mathcal O(z) \ri) z^{K\sigma_3}.
\label{outpostouter}
\ee

From the expression of  $\Psi_K$ (\ref{outerRHP4}) one can also write $[\A_K,\B_K]$  as follows
\be
[\A_K,\B_K]=\Psi_K z^{-K\sigma_3}=:[\A_K(0),\B_K(0)]+[\A_K'(0),\B_K'(0)] z+{\cal O}(z^2),
\ee
where we write the first two terms in the expansion for future reference.  
\bea\label{explicit1}
&&[\A_K(0),\B_K(0)]=\frac{\le(\frac{b-a}{4}\ri)^{r\sigma_3}t_0^{K\sigma_3}}{\sqrt{1-t_0^{-2}}}\le[\begin{array}{cc}1&it_0^{-1}\\-it_0^{-1}&1\end{array}\ri]
t_0^{r\sigma_3}\le(\frac{b-a}{4ab}\ri)^{K\sigma_3},
\\
&&[\A_K'(0),\B_K'(0)]=\frac{\le(\frac{b-a}{4}\ri)^{r\sigma_3}t_0^{K\sigma_3}}{\sqrt{1-t_0^{-2}}}
\le[\begin{array}{cc}0&\frac{i}{t_0\sqrt{ab}}\\-\frac{i}{t_0\sqrt{ab}}&0\end{array}\ri]t_0^{r\sigma_3}\le(\frac{b-a}{4ab}\ri)^{K\sigma_3}+
\\&&\qquad\qquad\qquad\qquad+[\A_K(0),\B_K(0)]
\le(\frac{1}{(t_0^2-1)\sqrt{ab})}{\bf 1}-\frac{r}{\sqrt{ab}}\sigma_3+\frac{K(b+a)}{2ab}\sigma_3 \ri).
\eea

{
\br
In \cite{MoBirth, Claeys} the authors used a scaling limit by removing the charges corresponding to the new zeroes by adding a point-wise charge with the same total mass. This amounts to multiplying the jump by  a factor $(x-\xi_0)^{2\varkappa}$ (in our notation). As a drawback they need to cure the non--constant jump residual after the lens-opening by introducing a scalar function $D(x)$ (Sz\"ego function) solving a new (scalar) RHP on the cut. In our case this scalar function is ``built-in'' the outer parametrix and corresponds to the term $\le(\frac{t - 1/t_0}{t- t_0}\ri)^K$. As a result the outer parametrices are at first sight of different nature, but -as it should- all terms can be put in correspondence  in the three approaches. 
\er
}

\subsection {Parametrix near the simple turning points}
\begin{wrapfigure}{r}{0.3\textwidth}
\resizebox{0.3\textwidth}{!}{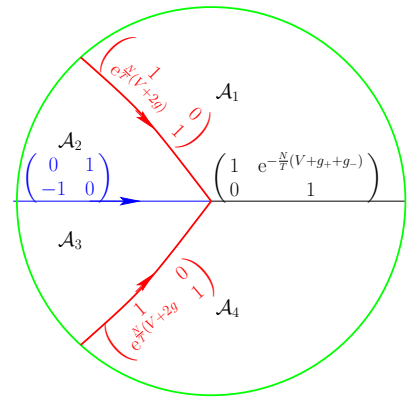}
\caption{The jumps of the exact solution $\wt Y$ near the soft-edge.}
\label{softedgeRHP}
\end{wrapfigure}

This part is essentially identical to the established results in \cite{Deift, DKMVZ}.
We consider only the right turning point at $z=b$; the method works for the other one.  Inside the green disk we solve the exact RHP.

For simplicity of exposition we consider the case where the turning point is simple, namely
\be \varphi(z)=\frac12 V(z)-g(z) = C(z-b)^{\frac 32} (1+ \mathcal O(z-b)),
\ee
for $z$ not on the cut. 
[We will indicate the trivial modifications needed in case of nonregular behavior later.]

We define a local coordinate by
\be
\frac 23\xi^{\frac 32} := \frac N T \varphi(z),
\ee
where the determination of the root is such that the (blue) cut is mapped to $\R_-$ of $\zeta$-plane.

We then introduce the standard Airy parametrix $\mathcal A^0(\xi)$
\cite{Deift, DKMVZ} as the piecewise defined matrix  $\mathcal A^0_j$
(see Figure \ref{softedgeRHP}) constructed in terms of the Airy function $\mbox{Ai}(x)$ as follows

\bea
\mathcal A^0_j(\xi)&\&:= \sqrt{2\pi} {\rm e}^{-\frac {i \pi}4}
 \le\{
\begin{array}{lc}\pmatrix{y_0 & -y_2\cr
y_0'& - y_2'}{\rm e}^{\frac 23\xi^{\frac 32}\sigma_3} & j=1\\[15pt]
\pmatrix{-y_1 & -y_2\cr
-y_1'& - y_2}{\rm e}^{\frac 23\xi^{\frac 32}\sigma_3} & j=2\\[15pt]
\pmatrix{-y_2 & y_1\cr
-y_2'&  y_1'}{\rm e}^{\frac 23\xi^{\frac 32}\sigma_3} & j=3\\[15pt]
\pmatrix{y_0 & y_1\cr
y_0'& y_1'}{\rm e}^{\frac 23\xi^{\frac 32}\sigma_3} & j=4
\end{array}
\ri.,
\eea

where we have used the definitions,
\be y_j:= \omega^j {\rm Ai}(\omega^j \xi),\ \ \ j=0,1,2,\ \ \omega = {\rm e}^{2i\pi/3}.\ee
Each of the above has the following uniform asymptotic behavior near $\xi=\infty$.
\be
\mathcal A^{0}(\xi) \sim
\xi ^{-\frac {\sigma_3}4}\frac 1 {\sqrt{2}} \pmatrix{ 1 &  1\cr -1 & 1} {\rm e}^{\frac {-i\pi \sigma_3}4 }
({\bf 1} +\mathcal O(\xi^{-3/2})).\label{formal}
\ee
This matrix has jumps on the rays (in the $\xi$--plane)  as indicated in Fig. \ref{Airy0} below.
%

\begin{wrapfigure}[16]{l}{0.3\textwidth}
\resizebox{0.3\textwidth}{!}{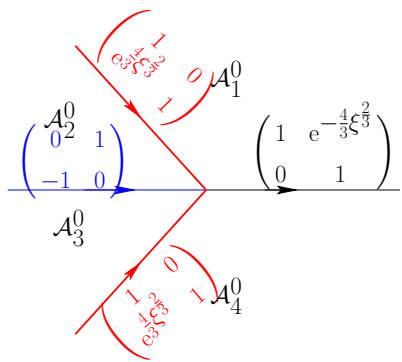}
\caption{The usual jumps for the local parametrix}
\label{Airy0}
\end{wrapfigure}

Thus, the final form of the local parametrix is simply
\be
\mathcal A_j(\xi):=
\overbrace{
{{\rm e}^{\frac {i\pi \sigma_3}4 }
\frac 1{\sqrt {2}} \le[
\begin{array}{cc}
1 &-1\\
1&1
\end{array}
\ri] \xi^{\frac {\sigma_3} 4} 
}
}^{:=F(z)}\mathcal A_j^{0}(\xi)
\ee

Such matrix has the properties
\begin{itemize}
\item It solves the exact jump conditions  of the RHP in Fig. \ref{softedgeRHP};
\item the prefactor $F(z)$ solves a RHP {\bf on the left}
\be
F(z)_+ =\le[
\begin{array}{cc}
0 &{- 1}\\
{1} & 0
\end{array}
\ri] F(z)_-\ ,\qquad \xi\in \R_-\ .
\ee
\item it behaves as $\mathcal A(\xi) = \1 + \mathcal O(N^{-1})$ uniformly on the boundary on account that $z^{K\sigma_3}$ is analytic and invertible with analytic inverse in the neighborhood.
The only point to raise is that $\Psi(z) \mathcal A(z)$ is {\bf bounded} inside the disk; indeed near the turning point we have
\be
\Psi = \mathcal O( (z-c)^{-\frac 14}),\quad F(z)= \mathcal O( (z-c)^{-\frac 14}).
\ee
Thus the product $\Psi(z)F(z)$ may at most have squareroot singularities: however, comparing the RHP that they solve, we see that the product is a single-valued matrix, thus must be analytic since at worst it may have singularities of type $(z-c)^{-\frac 12}$. Since $F(z)$ is solely responsible for the singularities arising in the local parametrix at $z=b$, this proves the assertion.
\end{itemize}
\br
If the turning points are non-regular then we should use the local parametrices for the usual problem as described in \cite{DKMVZ}. None of the above considerations (except for the bounds in the error terms) is significantly modified.
\er

\subsection{The local parametrix at the outpost}
At the outpost the effective potential behaves as $\varphi(z)=\frac {V(z)}2-g(z) + \frac \ell 2\simeq T  C_0z^{2\nu+2}$ with $C_0>0$.  We define a new conformal parameter $\tilde z$ as follows:
\be
\tilde z := C_0^{-\gamma}\frac 1 T \varphi(z)^{\gamma}=z+{\cal O}(z^2)\ee
We define $\mathbb D$ to be a finite open neighborhood around $z=0$ that maps univalently to a disk centered at $\tilde z=0$.
We also define the local coordinate
\be\label{zetadefine}\zeta:=C_0^\gamma N^{\gamma}\tilde z.\ee
The RHP satisfied by the {\em local parametrix} $R$ is as follows. (We will use $R$ with a subscript or a decoration such as in $\tilde R_K$, to indicate a specific solution to the RHP.) 
\bea
R_+ &\&= R_- \le[
\begin{array}{cc}
1 & {\rm e}^{-\zeta^{2\nu+2}} N^{2 \varkappa \gamma }\\
0&1
\end{array}
\ri],\quad\zeta\in \R,
\label{Hjump}
\\
R &\& \sim  \zeta^{-K \sigma_3} \mathcal O(1)\ ,\qquad \zeta \to 0\cr
R &\&\sim \1 +\mathcal O(N^{-\epsilon})
\ ,\qquad z \in \partial{\mathbb D}.
\label{boundary}
\label{newHRHP}
\eea
Here $\epsilon$ is some positive number that will be determined in the subsequent analysis.   Increasing $\epsilon$ leads to a better asymptotics.  In addition to the above conditions, we also require that $\Psi R$ is analytic in ${\mathbb D}$, which implies that $\det R=1$.

In section \ref{section_outerparametrix}, we obtain an outer parametrix $\Psi_K$ which has the pole behavior of order $K$ at the outpost.
We first look for the corresponding local parametrix, which we will call $R_K$.

\subsubsection{Case $\varkappa<0$}
We first observe that if $\varkappa<0$, then the solution is immediately written as\footnote{In fact all the results hold for $\varkappa<1/2$ as well.}
\be
R_0:= \le[
\begin{array}{cc}
1 &  \ds\frac {N^{2\varkappa \gamma } }{2i\pi} \int_\R \frac {{ \rm e}^{-\xi^{2\nu+2}}\d \xi}{\xi-\zeta}\\[20pt]
0&1
\end{array}
\ri]
\ee
On $\partial\D$ we have
\be
R\sim \1 + \mathcal O (N^{2\gamma\varkappa-\gamma})\ .
\ee
This situation is ``trivial'' from the point of view of the asymptotics and hence we will only focus on the case $\varkappa>0$ in the following.

\br
In fact we observe from the explicit solution that if $\varkappa<-2\nu-2$ then we may simply use the identity matrix for the local parametrix, thus committing an error smaller than $\mathcal O(N^{-1})$ which anyway arises on the boundary of the other turning points.
\er

\subsubsection {Case $\varkappa\geq 1/2$}
We recognize in (\ref{Hjump}) the Riemann--Hilbert problem of the orthogonal polynomials for the weight ${\rm e}^{-\zeta^{2\nu+2}} \d \zeta$. Specifically, if we denote by $P^{(\nu)}_\ell(\zeta)$ the monic orthogonal polynomials that satisfy
\be
\int_\R P^{(\nu)}_\ell(\xi) P^{(\nu)}_{\ell'}(\xi) {\rm e}^{-\xi^{2\nu+2}}\d \xi = \eta_{\ell} \delta_{\ell\ell'}\ ,\qquad \eta_\ell>0
\ee
then the solution of the RHP (\ref{Hjump}), (\ref{newHRHP}) is simply given by
\begin{equation}\label{RK}
R_K=\tilde z^{-K\sigma_3}H_K(\zeta),
\end{equation}
where
\bea
H_K(\zeta)&\&:=C_0^{-\gamma K\sigma_3}N^{\gamma\delta\sigma_3}\le[
\begin{array}{cc}
\ds
P_K^{(\nu)}(\zeta) & \ds \frac 1{2i\pi} \int_\R \frac{P_K^{(\nu)}(s){\rm e}^{-s^{2\nu+2}}\d s}{s-\zeta}
\\
\ds \frac {-2i\pi}{\eta_{K-1}}P_{K-1}^{(\nu)}(\zeta) & \ds\frac {-1}{\eta_{K-1}}  \int_\R \frac{P_{K-1}^{(\nu)}(s){\rm e}^{-s^{2\nu+2}}\d\xi}{s-\zeta}
\end{array} 
\ri] N ^{-K\gamma \sigma_3}N^{(K-\varkappa )\gamma\sigma_3}\cr
=&\& 
 C_0^{-\varkappa\gamma \sigma_3}\le[
\begin{array}{cc}
\ds
\Gamma ^{-K} P_K^{(\nu)}(\zeta) & \ds \frac { \Gamma^{2\varkappa  - K}} {2i\pi} \int_\R \frac{P_K^{(\nu)}(s){\rm e}^{-s^{2\nu+2}}\d s }{s-\zeta}\\
\ds
\frac {-2i\pi \Gamma^{K-2\varkappa } }{\eta_{K-1}}P_{K-1}^{(\nu)}(\zeta) & \ds\frac {-\Gamma^{K} }{\eta_{K-1}}  \int_\R \frac{P_{K-1}^{(\nu)}(s){\rm e}^{-s^{2\nu+2}}\d s}{s-\zeta }
\end{array} 
\ri]C_0^{\varkappa\gamma \sigma_3} \ ,\qquad \Gamma:=(C_0 N)^{\gamma}
\eea
It is crucial to point out here that the right multiplier $N^{-\gamma\varkappa \sigma_3}$  is needed to satisfy the correct jump relations (\ref{Hjump}), while the left multiplier $C_0^{-\gamma K\sigma_3}N^{\gamma\delta \sigma_3}$ is needed to restore the boundary condition (\ref{boundary}).
One can satisfy the boundary condition (only) by choosing $K$ as the closest integer of $\varkappa$.
Defining $\delta := \varkappa -K\in \le(-\frac 1 2 , \frac 12 \ri)$,  we obtain the following estimate holding {\bf uniformly on the boundary}.
\be
R_K= \1 + \mathcal O(N^{-\gamma+2|\delta|\gamma}),\quad z\in \pa\D\label{errorestimate2}.
\ee
Lastly, $\Psi_K R_K$ is analytic in ${\mathbb D}$, because $z^{-K\sigma_3}$ factor in (\ref{RK}) cancels out the singularity of $\Psi_K$ (\ref{ABK}).

The important observation is that if $\varkappa \in \frac 1 2 + \Z$ then the error term in (\ref{errorestimate2}) does not tend to zero (it is $\mathcal O(1)$).
It is understandable as these values separate regimes where the value of $K$ jumps by one unit and the whole strong asymptotic must changes its form.  A similar problem arose in \cite{EynardBirth}. In section \ref{sectimproved} we will overcome this obstacle.

\br
The orthogonal polynomials we are using here are a particular case of the so--called {\em Freud Orthogonal Polynomials} \cite{Freud}.
\er


%
%
\subsection{Asymptotic solution for $\wt Y$ and error term}
Collecting the results of the above analysis we have the following asymptotic solution for $\tilde Y$.
\bea
\tilde Y_{\mbox{as}}:=\le\{
\begin{array}{cl}
\ds\Psi_K(z) & \hbox {outside of the disks }\\[5pt]
\ds \Psi_K(z) R_K &  z\in\D\\[5pt]
\ds \Psi_K(z) \mathcal A& \hbox { inside the disks around the turning points}
\end{array}
\ri.
\eea

To find the error term we define the {\em error matrix} as follows.
\bea
\mathcal E(z):=\tilde Y\tilde Y^{-1}_{\mbox{as}}=\le\{
\begin{array}{cl}
\ds\tilde Y\Psi^{-1}_K(z) & \hbox {outside of the disks }\\[5pt]
\ds \tilde YR^{-1}_K\Psi^{-1}_K(z)  &  z\in\D\\[5pt]
\ds \tilde Y\mathcal A^{-1}\Psi^{-1}_K(z) & \hbox { inside the disks around the turning points}
\end{array}
\ri.
\eea

\begin{wrapfigure}{r}{0.7\textwidth}
\resizebox{0.7\textwidth}{!}{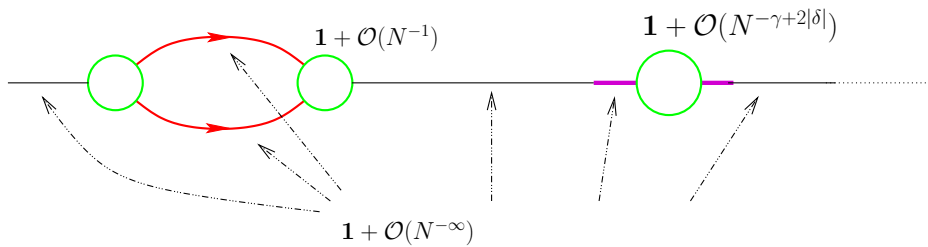}
\caption{\label{fig_Residual} The jumps of ${\mathcal E}$.}
\end{wrapfigure}

The error matrix solves the residual RHP with the jump matrices as shown in the figure \ref{fig_Residual}.
It follows from the construction that there is no jump inside the green disks, and on the cut.
On the disks with airy parametrix and on $\partial \D$, the jumps converge to the identity with uniform error bounds in $L^2\cap L^\infty$.
We evaluate these error bounds now.
 
The jump of ${\mathcal E}$ on $\partial \D$ is evaluated as follows.
\be
{\mathcal E_-^{-1} \mathcal E_+ = \Psi_K R^{-1}_K \Psi_K^{-1}={\bf 1}+{\cal O}(N^{-\gamma+2\gamma|\delta|}).}
\ee
The jumps of ${\mathcal E}$ on the disks of Airy parametrix is similarly evaluated as follows.
\be
{\mathcal E_- ^{-1}\mathcal E_+ = \Psi_K {\cal A}^{-1} \Psi_K^{-1}={\bf 1}+{\cal O}(N^{-1}).}
\ee

Therefore, the error matrix has jump matrices that are uniformly close to the identity with the error bound of ${\cal O}(N^{-\gamma+2\gamma|\delta|})$ in $L^2\cap L^\infty$.
A well-known theorem \cite{Deift} guarantees that error matrix itself is bounded by the same error bound, i.e. ${\mathcal E}={\bf 1}+{\cal O}(N^{-\gamma+2\gamma|\delta|})$.  This gives the following error term for the (strong) asymptotics of $\tilde Y$.
\be
\tilde Y=\tilde Y_{\mbox{as}}({\bf 1}+{\cal O}(N^{-\gamma+2\gamma|\delta|})).
\ee

\subsection{Necessity of improved approximation: a nonlinear Stokes phenomenon in $\varkappa$}
From the above estimate of the error term it appears that our global parametrix does a poor approximation if $\varkappa$ is not exactly an integer, and it is no approximation at all if $\varkappa\in \frac 12 + \Z$ since the error term is not vanishingly small. These transition points are the equivalent of the Stokes' lines in the standard theory of asymptotics of ODEs, where two solutions of the same RHP become of the same magnitude whereas off the line one is recessive and the other dominant.

The phenomenon is similar here: for $\varkappa \in \le(K-\frac 12, K+\frac 12 \ri)$ the dominant solution is the one we have constructed with $K$, where for $\varkappa\in \le(K+\frac 12, K+\frac 32\ri)$ it is the ``next'' with $K+1$.

For $\varkappa = K+\frac 12$ it is to be expected that both solution are of the same magnitude in a suitable sense and hence a ``linear combination'' should be sought for.

We show how to do this in the next section and we will construct hence a parametrix including the subleading term in the $N^{-\gamma}$ expansion so as to have a uniform  approximation to within $\mathcal O(N^{-2\gamma})$.
\section{Asymptotic solution for $\varkappa \in \mathbb R$ up to ${\cal O}(N^{-2\gamma})$}
\label{sectimproved}
\subsection{A short digression on Schlesinger transformations}
We start by observing that $\Psi_K$ and $\Psi_{K+1}$ are related by a Schlesinger transformation, as well as the local parametrices $R_K, R_{K+1}$. In particular, we consider the ``raising'' Schlesinger transformation that raises the order of poles by one
\be\label{raising}
\Psi_{K+1}(z) =  \le(\1 + \frac {S_K} z\ri)\Psi_K(z)\ ,
\ee
for a $z$-independent matrix $S_K$.
To show this, it is sufficient to observe that $\Psi_{K+1}\Psi_K^{-1}$ has no jumps, behaves as the identity at infinity and has at most a simple pole at the outpost. The formula follows immediately from Liouville's theorem.
Note also that $S_K$ is a {\bf nilpotent matrix} as follows immediately from the fact that $\det(\1 + S_K/z)\equiv 1$.

For future reference we compute $S_K$.  Let us write the outer parametrix $\Psi_K$ as 
\be\label{ABK}
\Psi_K = [\A_K, \B_K] z^{K\sigma_3}
,
\ee
where $[\A_K, \B_K]$ {is a $2\times 2$ matrix holomorphic at $z=0$}.  We also require $\det[\A_K,\B_K] \equiv 1$ to have $\det\Psi_K=1$.

Let us compute $S_K$ in the  raising Schlesinger transformation; the condition that determines $S_K$ is that 
\be
\A_K(z) +\frac 1 z S_K\A_K(z) = \mathcal O(z)
\ee 
or, equivalently,
\be
\le\{
\begin{array}{cc}
S_K \A_K(0)= 0 \\[5pt]
\A_K(0) + S_K\A_K'(0)=0
\end{array}
\ri.\ \Rightarrow\quad
S_K{\bf V}= \frac {\det[{\bf V},\A_K(0)]}{\det[\A_K(0), \A_K'(0)]}\A_K(0) \ ,\quad\forall
{\bf V}\in \C^2.
\ee

Similarly the inverse ``descending'' transform
\be
\le (\1 + \frac {{\wt S}_K}z\ri) \Psi_{K} = \Psi_{K-1},
\ee
requires the analyticity condition
\be
\B_K (z)+ \frac 1z \wt S_K \B_K(z)  = \mathcal O(z).
\ee
This determines $\wt S_K$ as
\be
\wt S_K{\bf V} =  \frac {\det[{\bf V}, \B_K(0)]}{\det[\B_K(0), \B'_K(0)]}\B_K(0),
\ee
for an arbitrary vector ${\bf V}$.
It is also important to note that, since we have explicitly constructed the sequence of $\{\Psi_K|K=0,1,\cdots\}$ in (\ref{outerK}), the Schlesinger transformations between them must exist, and hence
\be
\det[\B_K(0), \B'_K(0)] \neq 0 \neq \det[\A_K(0), \A'_K(0)]\ ,\quad\forall K\in \Z\label{nontau}.
\ee

Now let us consider the similar transformation for the local parametrix.
Previously in (\ref{RK}) we have constructed the local parametrix $R_K=\wt z^{-K\sigma_3}H_K$ where
\begin{equation}\label{HK}
H_K
\simeq\le[\begin{array}{cc}1&-u_K/\tilde z\\-\ell_{K-1}/\tilde z&1\end{array}\ri]\tilde z^{K\sigma_3}.
\end{equation}
All the components of the matrix in RHS have the multiplicative error bounds of $1+{\cal O} (N^{-2\gamma}\wt z^{-2})$.  
The two variables $u_K$ and $\ell_{K-1}$ are given by
\be\label{mumutilde}
u_K=\frac{\eta_K}{2 i\pi }\frac{N^{2\gamma\delta-\gamma}}{C_0^{2\gamma K+\gamma}},
\qquad
\ell_{K-1}=\frac{2i\pi}{\eta_{K-1}} \frac{ C_0^{2\gamma K-\gamma}}{N^{2\gamma\delta+\gamma}}.\ee
Note the interesting relation $u_K\ell_K=1$, which will be essential for the consistency of our solution.

From a similar argument used for deriving (\ref{raising}) using Liouville theorem, one obtains the following relation
\be\label{Hraising}
H_{K+1}= \le[
\begin{array}{cc}
\ds \wt z  &u_K\\
-\ell_K& 0
\end{array}
\ri] H_{K}.
\ee
This transform can also be derived from the three-term recurrence relation
\be
\zeta P_K(\zeta) = P_{K+1}(\zeta) + \frac {\eta_{K}}{\eta_{K-1}} P_{K-1}(\zeta).
\ee
The transformation matrix in (\ref{Hraising}) is LDU decomposed as follows
\be
\le[
\begin{array}{cc}
\ds \wt z  &u_K\\
-\ell_K& 0\end{array}\ri]
= L_K^{-1} \wt z^{\sigma_3} U_K, \qquad 
L_K:=\le[\begin{array}{cc}1 &0\\\ell_K/\wt z& 1\end{array}\ri],\quad
U_K:=\le[\begin{array}{cc}1 &u_K/\wt z\\0& 1\end{array}\ri].
\label{facto}
\ee 
From this we may view the matrices $L_K, U_K$ as the ``two halves'' of the transformation that raises the order of the Freud's OP by one.
Therefore, these objects will appear for the ``half-raising" transform, or more generally for the continuous transform that is parametrized by $\delta$ in the next section.

\subsection{Improved parametrices}

Here we improve both the outer parametrix $\Psi_K$ and the local parametrix $R_K$ to produce a legitimate asymptotics at half-integer $\varkappa$ and to produce a better asymptotics for all $\varkappa$.
Especially, we will use certain transformations that resemble the Schlesinger transformation discussed in the previous section.

Looking back at the error analysis, the dominant error (\ref{errorestimate2}) originates from the off-diagonal term of $H_K\wt z^{K\sigma_3}$ (\ref{HK}), especially from the terms $u/\wt z$ and $\ell/\wt z$.
 
A natural way to correct this problem is to define a new local parametrix,\footnote{Our notation is not exactly consistent, since $R_\varkappa\neq R_K$ when $\varkappa=K$.  However, we believe that this will not make any confusion.}
\be\label{advertise}
R_{\varkappa}:=
\tilde z^{-K\sigma_3}U_K L_{K-1}
H_K={\bf 1}+{\cal O}(N^{-2\gamma}),\ \ z\in \pa\D
\ee
so as to cancel out the leading off-diagonal terms.
We get an improved error bound which is independent of $\varkappa$.

We may also change the order of $L_{K-1}$ and $U_K$ in the above definition;
although the two matrices $L_{K-1}$ and $U_K$ do not commute, the non-commutativity is within the error bound of ${\cal O}(N^{-2\gamma})$ and, therefore, the order does not make any difference in the asymptotics.

Given the local parametrix $R_\varkappa$, we will find the corresponding {\em improved} outer parametrix $\Psi_\varkappa$.
It is notable that $R_\varkappa$ is constructed out of $L_{K-1}$ and $U_K$, which appear in the raising transform (\ref{facto}).
This suggests that the appropriate outer parametrix $\Psi_{\varkappa}$ may come from a ``partial 2-step Schlesinger'' transformation of $\Psi_K$, such as $\Psi_\varkappa = (\1 + F_1/z + F_2/z^2) \Psi_K$.
$F_1$ and $F_2$ are then determined by the analyticity of $\Psi_\varkappa R_\varkappa$ at $z=0$ or, equivalently, 
\be\label{F1F2}
\le(\1 + \frac {F_1} z +\frac {F_2}{z^2} \ri) \Psi_K  \wt z^{-K\sigma_3}
U_KL_{K-1}  = \mathcal O(1), \quad\mbox{in}~\D.
\ee
Clearly the issue  is to remove the possible poles at $z=0$, and the problem is addressed in the next section.
\subsection{Improved outer parametrix}

The improved outer parametrix, that we will denote by $\Psi_\varkappa$\footnote{Note --once more-- that $\Psi_\varkappa\neq \Psi_K$ when $\varkappa=K$ as for $R_\varkappa$.  However, we believe that this will not make any confusion since we are now constructing a refinement of the previous setting.}, must satisfy the analyticity condition (\ref{F1F2}).
$\Psi_\varkappa$ is uniquely solved by the condition (\ref{F1F2}), and we will solve it in two steps by writing $\Psi_\varkappa$ by first writing
\be\Psi_\varkappa=\le(\1 + \frac {G}{z}\ri)\le(\1 + \frac {F}{z}\ri)\Psi_K.\ee
Remembering $\Psi_K=[\A_K,\B_K]z^{K\sigma_3}$ (\ref{ABK}),
we can rewrite $\Psi_\varkappa R_\varkappa$ as follows.
\be \label{zwtz}
\Psi_\varkappa R_\varkappa=\le(\1 + \frac {G}{z}\ri)\le(\1 + \frac {F}{z}\ri)[\wt\A_K,\wt\B_K]\le[\begin{array}{cc}1&u_K/z\\ 0&1\end{array}\ri]\le[\begin{array}{cc}1&0\\ \ell_{K-1}/z&1\end{array}\ri] \wt H_K.
\ee
Note that the terms $u/z$ and $\ell/z$  depend on $z$ and {\em not} on $\wt z$.  Accordingly we have the following definitions.
\be
[\wt\A_K,\wt\B_K]:=[\A_K,\B_K](z/\wt z)^{K\sigma_3}\le[\begin{array}{cc}1&\frac{u_K}{\wt z}-\frac{u_K}{z}\\0&1\end{array}\ri],
\ee
which is analytic at $z=0$, and
\be
\wt H_K:=\le[\begin{array}{cc}1&0\\\frac{\ell_{K-1}}{\wt z}-\frac{\ell_{K-1}}{z}&1\end{array}\ri] H_K.
\ee

The first step is to determine $F$ by {\em imposing} the analyticity condition: (and we define another notation as below)
\be\label{ABtilde}
[\hat\A_K, \hat\B_K]:=\le(\1 + \frac {F}{z}\ri) [\wt\A_K, \wt\B_K] 
\le[\begin{array}{cc}1&u_K/z\\ 0&1\end{array}\ri]={\cal O}(1).
\ee
Assuming we solved the above, the second step is to determine $G$ by imposing  the new analyticity condition:
\be\le(\1 + \frac {G}{z}\ri) [\hat\A_K,\hat\B_K] 
\le[\begin{array}{cc}1&0\\ \ell_{K-1}/z&1\end{array}\ri]={\cal O}(1).
\ee
The above two analyticity conditions uniquely determine $F$ and $G$, which can be written as follows using an arbitrary vector ${\bf V}$.
\be\label{FG}
F {\bf V}= \frac {u_K\det[{\bf V},\wt\A_K(0)]}{1+u_K\det[\wt\A_K(0),\wt\A'_K(0)]} \wt\A_K(0),\quad
G {\bf V}= \frac {-\ell_{K-1}\det[{\bf V},\hat\B_K(0)]}{1+\ell_{K-1}\det[\hat\B_K'(0),\hat\B_K(0)]} \hat\B_K(0).
\ee
Here we have used the fact that $\det[\wt\A_K(0),\wt\B_K(0)]=\det[\hat\A_K(0),\hat\B_K(0)]=1$.  

From the above formulae it is easily noticed that $F\sim{\cal O}(u_K)$ and $G\sim{\cal O}(\ell_{K-1})$.  This immediately tells that $[\hat \A_K,\hat \B_K]=[\wt \A_K,\wt \B_K]+{\cal O}(u_K)$ from the definition (\ref{ABtilde}).  This is useful to know because now we can change all the $\hat\B_K$'s into $\wt\B_K$'s in the above equation (\ref{FG}) while keeping the error under ${\cal O}(u_K\ell_{K-1})={\cal O}(N^{-2\gamma})$.
Since our asymptotic solution will have an error bound of ${\cal O}(N^{-2\gamma})$ any term of that order or lower is meaningless.
In addition, it is useful to observe the following facts for a further simplification.
\bea
&&\wt\A_K(0)=\A_K(0),\qquad \det[\wt\A_K(0),\wt\A'_K(0)]=\det[\A_K(0),\A_K'(0)],
\\&&\wt\B_K(0)=\B_K(0),\qquad \det[\wt\B_K(0),\wt\B'_K(0)]=\det[\B_K(0),\B_K'(0)]+{\cal O}(u_K).
\eea

As a result we obtain (the first column of) $\Psi_\varkappa$ up to ${\cal O}(N^{-2\gamma})$ as follows.
\bea\label{firstcol}
\Psi_\varkappa\big|_{(1)}&\&:= \le(\1 + \frac {G}{z}\ri)\le(\1 + \frac {F}{z}\ri)\A_K z^K
\\\nonumber&\&\hspace{-20pt}\simeq \A_K z^K+ \frac {u_K\det[\A_K,\A_K(0)]}{1+u_K\det[\A_K(0),\A'_K(0)]} \A_K(0) z^{K-1}
-\frac {\ell_{K-1}\det[\A_K,\B_K(0)]}{1+\ell_{K-1}\det[\B_K'(0),\B_K(0)]} \B_K(0) z^{K-1}+{\cal O}(N^{-2\gamma}).
\eea
The second column is analogously given.
In the last formula, when $\delta\neq\pm 1/2$, the first term provides the leading term, which becomes the strong asymptotics for OPs away from the outpost.  Therefore, the strong asymptotics is $\delta$ independent.

The subleading term is either from the second or the third term depending on the value of $\delta$; to decide, we must recall the scaling behaviors of $u_K\sim N^{\gamma(2\delta-1)}$ and $\ell_{K-1}\sim N^{-\gamma(2\delta+1)}$ (\ref{mumutilde}).

At both $\delta=\pm 1/2$, the leading strong asymptotics changes.
At $\delta=1/2$, $u_K$ is no longer scaling with $N$ and, therefore, the second term also contributes to the leading asymptotic behavior.
\be
\Psi_{K+1/2}\big|_{(1)}=\A_K z^K+ \frac {u_K\det[\A_K,\A_K(0)]}{1+u_K\det[\A_K(0),\A'_K(0)]} \A_K(0) z^{K-1}+{\cal O}(N^{-2\gamma}).
\ee
One also observes that the number of roots at the outpost is still $K$ as for $-1/2<\delta<1/2$.

At $\delta=-1/2$, $\ell_{K-1}$ is the non-scaling parameter  (see (\ref{mumutilde})) and, therefore, the third term contributes to the leading asymptotic behavior.  
\be
\Psi_{K-1/2}\big|_{(1)}=\A_K z^K-\frac {\ell_{K-1}\det[\A_K,\B_K(0)]}{1+\ell_{K-1}\det[\B_K'(0),\B_K(0)]} z^{K-1}\B_K(0) z^{K-1}+{\cal O}(N^{-2\gamma}).
\ee
In this case the number of roots is $K-1$; one less than what it is for $-1/2<\delta<1/2$.  The location of the missing root 
can also be found using the above expression.

To summarize, 
\begin{itemize}
\item[i)] At half-integer $\varkappa$, the number of roots at the outpost is given by the closest integer that is {\em smaller} than $\varkappa$.  
\item[ii)] At half-integer $\varkappa$, the outer parametrix $\Psi_\varkappa$ cannot be obtained by approaching from {\em either} side of $\varkappa$.
\end{itemize}  

\subsection{Roots at the outpost}
\label{sectroots}

So far, we have described the strong asymptotics away from the outpost.
Now we turn our attention to the inside of the disk ${\mathbb D}$ to look closely at the locations of the roots at the outpost.
To this purpose, we evaluate $\Psi_{\varkappa}R_\varkappa$ up to ${\cal O}(N^{-2\gamma})$ using (\ref{zwtz}).  It is a straightforward but long calculation if one tries to obtain the full asymptotics up to ${\cal O}(N^{-2\gamma})$.  
Instead, here we will obtain only the leading term and the subleading term.
(We write the full asymptotics in Appendix \ref{fullasymptotics} for reference.)

Looking at the first column of $\Psi_{\varkappa}R_\varkappa$ one gets
\bea\label{11delta+}
\left(N^{\gamma}C_0^{\gamma}\right)^K \Psi_\varkappa R_\varkappa\big|_{(1)}=\Big(\A_K(0)+{\cal O}(N^{-\gamma+2|\delta|\gamma})\Big)P^{(\nu)}_K(\zeta)
\\-\Big(\B_K(0)+{\cal O}(N^{-\gamma+2|\delta|\gamma})\Big)C_0^\gamma N^\gamma \ell_{K-1}P^{(\nu)}_{K-1}(\zeta).
\eea
Note that the above is the sum of two Freud's OPs.
One may object that, say for a positive $\delta$, the second term is within the error of the first term, and cannot contribute as a subleading term.   (For a negative $\delta$ it is the first term that provides the subleading correction.) A closer look shows however  that  both error terms are $\zeta$-independent up to ${\cal O}(N^{-\gamma})$ (which is not difficult to see from the general structure of the formula).  So the error terms only change the coefficients of the two polynomials up to ${\cal O}(N^{-\gamma})$.

\begin{wrapfigure}{r}{0.4\textwidth}
\includegraphics[width=0.4\textwidth]{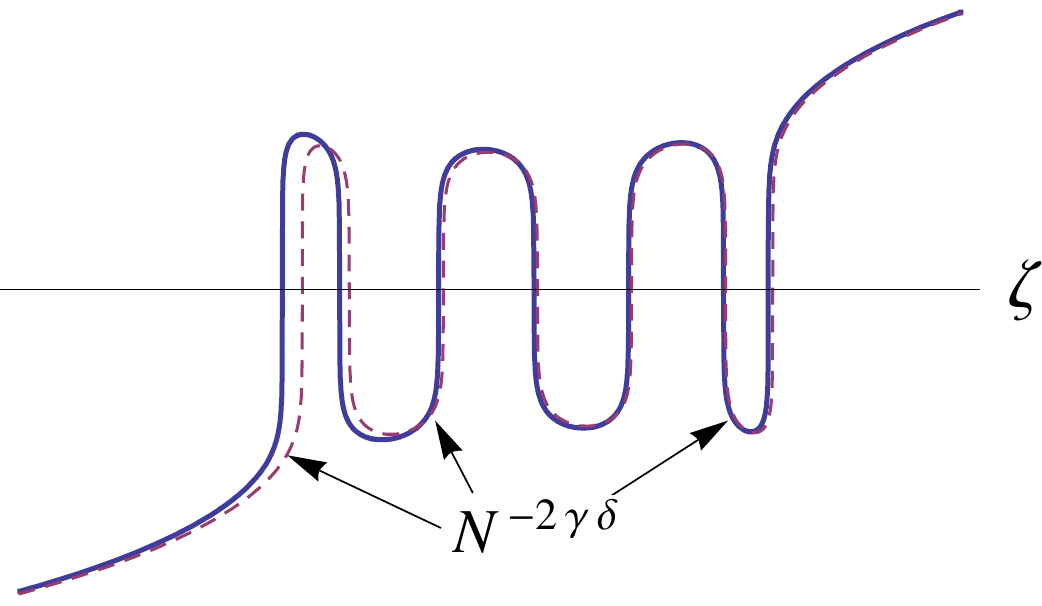}
\caption{For $\delta>0$, a schematic plot of polynomial at the outpost. The dashed line is the plot of $P_K^{(\nu)}(\zeta)$.  The deviation between the two plots is of the order $N^{-2\gamma\delta}$.\label{fig_delta+}}
\includegraphics[width=0.4\textwidth]{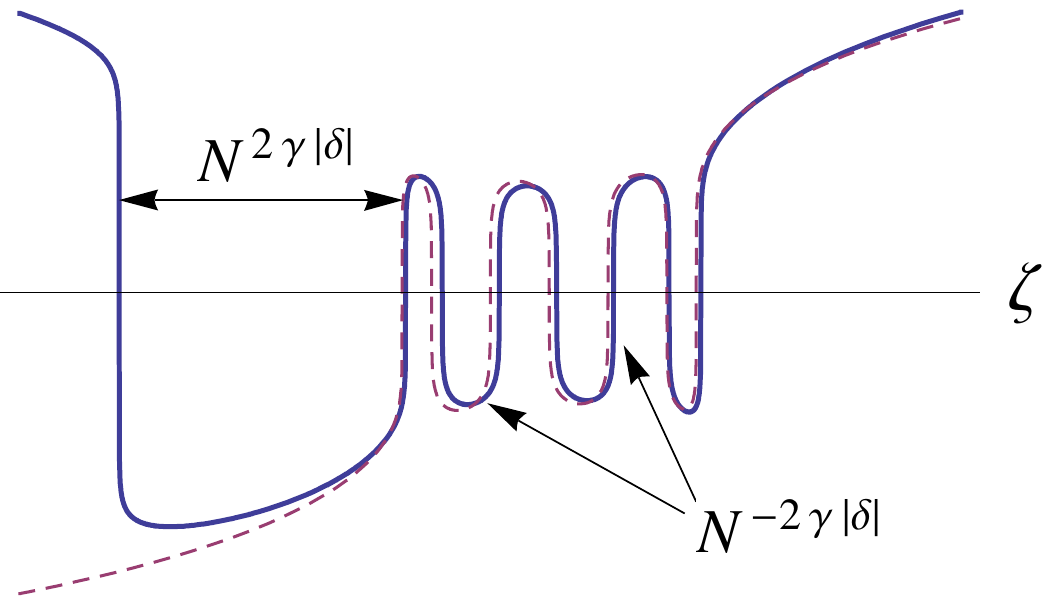}
\caption{For $\delta<0$, a schematic plot of polynomial at the outpost. The dashed line is the plot of $P_{K-1}^{(\nu)}(\zeta)$.  One of the zeros is found away from the rest.  In the $\tilde z$-plane, however, all the zeros converges to $\tilde z=0$.\label{fig_delta-} }
\end{wrapfigure}

From (\ref{11delta+}) we now identify the asymptotic locations of the roots at the outpost.
Let us first consider a positive $\delta$.  The roots are determined by the zeros of $P_K^{(\nu)}(\zeta)$ at the leading order.  
According to the explicit value of $[\A_K,\B_K]$ (\ref{explicit1}), the subleading term $N^{-2\delta\gamma}P_{K-1}^{(\nu)}(\zeta)$ contributes with a same sign as the leading term.

Due to the well-known interlacing property of the OPs, this means that all the roots are shifted from the zeros of $P_K^{(\nu)}(\zeta)$
by the amounts that scale as $N^{-2\delta\gamma}$ in $\zeta$-coordinate. 

In the figure \ref{fig_delta+} we show a schematic view of OPs and the roots where the real OP is shifted {(to the left)} from the leading asymptotics (Freud's OP; dashed line) by $O(N^{-2\gamma\delta})$. 

For  negative $\delta$ the leading asymptotics is now given by $P_{K-1}^{(\nu)}(\zeta)$ and therefore we only see $K-1$ roots to this order.  Also from the explicit value of $[\A_K,\B_K]$, the subleading term $N^{-2\gamma|\delta|}P_K^{(\nu)}(\zeta)$ contributes with the same sign as the leading term, which means that all the $K-1$ roots shift to the {right} by ${\cal O}(N^{2\delta\gamma})$.

Most interestingly there appears another root (let us call it ``the stray root") to the left of $K-1$ roots, distanced by $\sim N^{2\gamma|\delta|}$.
As in the schematic view (figure \ref{fig_delta-}) the stray root scales differently from all the other roots.
Though this root escapes to infinity in the $\zeta$-coordinate, it is actually converging to $z=0$, only much more slowly than the other roots.
It is also interesting to observe that, from the direction of the stray root, it seems to come from the main cut.
The approximate location of the stray root is given by
\begin{equation}\label{localstray}
\zeta_{\mbox{stray}}\sim \frac{C_0^{{2K\gamma} }}{N^{2\gamma\delta}}\frac{B_x(0)}{A_x(0)}\frac{2i\pi}{\eta_{K-1}}.
\end{equation}

Finally let us consider the cases $\delta=\pm1/2$.
For both cases we get an additional subleading term.
As the formulae are not particularly illuminating we present them in Appendix \ref{fullasymptotics}.

\subsection{Kernel at the outpost}

Questions regarding universality hinge on the behavior of the kernel for the correlators \cite{MehtaBook} in the scaling coordinate

In general the kernel is defined by 
\bea
K_n(z,z')&\& :=\frac{1}{h_{n-1}}\frac{p_n(z)p_{n-1}(z')-p_{n-1}(z)p_n(z')}{z-z'}\cr
&\& =\frac{1}{2i\pi}\frac{[Y^{-1}(z')Y(z)]_{21}}{z-z'}=\cr
&\&=\frac{1}{2i\pi}\frac{[\wt Y^{-1}(z')\wt Y(z)]_{21}}{z-z'}
{\rm e}^{-\frac NT\le(g(z)+g(z')+\ell\ri)}.
\eea
Using our asymptotics: $\wt Y\simeq \Psi_\varkappa R_\varkappa$, we evaluate the kernel in the local coordinate near the outpost
\bea
K_n(\zeta,\zeta')&\simeq&\frac{{\rm e}^{-\frac NT\le(g(z)+g(z')+\ell\ri)}}{2i\pi N^\gamma C_0^\gamma}\frac{\det\le[\Psi_\varkappa R_\varkappa(\zeta)\big|_{(1)},\Psi_\varkappa R_\varkappa(\zeta')\big|_{(1)}\ri]}{\zeta-\zeta'}.
\\&\simeq&-\frac{{\rm e}^{-\frac NT\le(g(z)+g(z')+\ell\ri)}}{N^{2\gamma\varkappa} (NC_0)^{\gamma}}\frac{P^{(\nu)}_K(\zeta)P^{(\nu)}_{K-1}(\zeta')-P^{(\nu)}_K(\zeta')P^{(\nu)}_{K-1}(\zeta)}{\eta_{K-1}(\zeta-\zeta')}\big(1+{\cal O}(N^{-\gamma+2\gamma|\delta|})\big).
\eea
Using the expression of $\Psi_\varkappa R_\varkappa$ in the appendix we can in principle obtain the kernel up to ${\cal O}(N^{-2\gamma})$ accuracy.  Here we only show the leading term.
The above approximation is valid  for $\varkappa\not\in \N+\frac 1 2$.
This is exactly the kernel for Freud's OPs, as we expect from the previous sections.

At $\varkappa=K+1/2$ we obtain a different kernel at the leading order.
\be\label{kernelplus}
K_n(\zeta',\zeta)=
-\frac{{\rm e}^{-\frac NT\le(g(z)+g(z')+\ell\ri)}}{N^{2\gamma\varkappa} (NC_0)^{\gamma}}
\left(\frac{P^{(\nu)}_K(\zeta)P^{(\nu)}_{K-1}(\zeta')-P^{(\nu)}_K(\zeta')P^{(\nu)}_{K-1}(\zeta)}{\eta_{K-1}(\zeta-\zeta')}+ \frac{\alpha_K}{\eta_{K}} P_K(\zeta')P_{K}(\zeta) \right)
\big(1+{\cal O}(N^{-\gamma})\big),
\ee
where the constant $\alpha_K$ is given by
\be
\alpha_K:=\frac{u_K \det[\A_K(0),\A_K'(0)]}{1+u_K\det[\A_K(0),\A_K'(0)]}.\ee

At $\varkappa=K-1/2$ we obtain the following kernel at the leading order.
\be
K_n(\zeta',\zeta)=
-\frac{{\rm e}^{-\frac NT\le(g(z)+g(z')+\ell\ri)}}{N^{2\gamma\varkappa} (NC_0)^{\gamma}}
\left(\frac{P^{(\nu)}_K(\zeta)P^{(\nu)}_{K-1}(\zeta')-P^{(\nu)}_K(\zeta')P^{(\nu)}_{K-1}(\zeta)}{\eta_{K-1}(\zeta-\zeta')}+ \frac{\beta_{K-1}}{\eta_{K}} P_{K-1}(\zeta')P_{K-1}(\zeta) \right)
\big(1+{\cal O}(N^{-\gamma})\big),
\ee
where the constant $\beta_{K-1}$ is given by
\be
\beta_{K-1}:= -\frac{\ell_{K-1}\det[\B_K'(0),\B_K(0)]}{1+\ell_{K-1}\det[\B_K'(0),\B_K(0)]}.\ee

\subsection{Some consistency checks}\label{section_hmmm}

So far we have seen that the various asymptotic properties are drastically changed when $\varkappa$ is a half integer.
For $\varkappa=K+1/2$ there are two ways to construct the asymptotics; one can apply a suitable (Schlesinger type) transformation starting from $\Psi_K$ or from $\Psi_{K+1}$.
In this section, the consistency of the two approaches will be proved in a completely general setting; without referring to the explicit global parametrix.

When $\delta=1/2$,  we can obtain the same error bound using only $U_K$ without $L_{K-1}$
\be
 R_{K+1/2}:=\wt z^{-K\sigma_3}U_KH_K.\label{above2}
 \ee
Correspondingly, the outer parametrix will then be written as 
\be \Psi_{K+1/2}:=\le({\bf 1}+\frac Fz\ri)\Psi_K,\ee
where $F$ has been explicitly solved for in (\ref{FG}).

The same state can be approached from $K+1$ by applying the ``half-descending" transformation.  In this spirit, the local parametrix may be written as follows. 
\be R^{(\mbox{new})}_{K+1/2}:=\wt z^{-(K+1)\sigma_3}L_{K}H_{K+1}.\ee
Consistency means that $R^{(\mbox{new})}_{K+1/2}=R_{K+1/2}$, which is an elementary consequence of the relations  (\ref{Hraising}) together with the ``factorization'' (\ref{facto}).   One realizes that $u_K\ell_K=1$ is the key identity.

We can then find the corresponding outer parametrix $\Psi^{(\mbox{new})}_{K+1/2}$ by demanding analyticity on $\Psi^{(\mbox{new})}_{K+1/2}R^{(\mbox{new})}_{K+1/2}$.
To prove the consistency of the outer parametrix we write
\be \Psi_{K+1/2}:=\le({\bf 1}+\frac {\wt F}z\ri)\Psi_{K+1}=\le({\bf 1}+\frac {\wt F}z\ri)\le({\bf 1}+\frac {S_K}z\ri) \Psi_{K},\ee
using the property of the Schlesinger transform.

Therefore, the proof of consistency amounts to the following identity:
\be\label{hmmm}
\le({\bf 1}+\frac {F}z\ri)=\le({\bf 1}+\frac {\wt F}z\ri)\le({\bf 1}+\frac {S_K}z\ri).\ee
To prove this, remember the following identity
\begin{equation}\label{2.16}
\left({\bf 1}+\frac{S_K}{z}\right)[\A_K,\B_K]=[\A_{K+1},\B_{K+1}]z^{\sigma_3}.
\end{equation}
Here we do not use $\wt{}$ above $\A$ and $\B$ as we are dealing with a generic case.
Then $F$ and $\wt F$ are defined by the following conditions.
\begin{equation}
\left(1+\frac{F}{z}\right)[\A_K,\B_K]
\left[\begin{array}{cc}1&u_K/z\\0&1\end{array}\right]=\mathcal O(1),
\end{equation}
\begin{equation}
\left(1+\frac{\wt F}{z}\right)[\A_{K+1},\B_{K+1}]
\left[\begin{array}{cc}1&0\\\ell_K/z&1\end{array}\right]=\mathcal O(1).
\end{equation}
To prove (\ref{hmmm}) we only need to show that the following quantity is analytic.
\begin{equation}
\left(1+\frac{\tilde F}{z}\right)\left(1+\frac{S_K}{z}\right)[\A_K,\B_K]
\left[\begin{array}{cc}1&u_K/z\\0&1\end{array}\right]=
{\cal O}(1)\left[\begin{array}{cc}1&0\\-\ell_K/z&1\end{array}\right]
z^{\sigma_3}\left[\begin{array}{cc}1&u_K/z\\0&1\end{array}\right].
\end{equation}
It is straightforward to see that the above does not have a pole (hence analytic), using the identity $u_K\ell_K=1$. 
This concludes that $\Psi^{(\mbox{new})}_{K+1/2}=\Psi_{K+1/2}$, as it should.

Now let us look at the kernel obtained in the previous section.
At $\varkappa=K+1/2$ the kernel has an additional term to the usual kernel from OPs.  To make sense of the additional term we recall the following general identity for OPs.
\be \frac{P_n(\zeta)P_{n-1}(\zeta')-P_n(\zeta')P_{n-1}(\zeta)}{\eta_{n-1}(\zeta-\zeta')}=\sum_{j=0}^{n-1}\frac{P_j(\zeta)P_j(\zeta')}{\eta_j}.
\ee
This tells us that the ``raising operation" for the kernel is to add a term of the form $\propto P_n(\zeta)P_n(\zeta')$, which happens to be the new term appearing in (\ref{kernelplus}).

As the kernel at $\varkappa=K+1/2$ can be approached from either $K$th kernel by adding $\alpha_KP_K(\zeta)P_K(\zeta')$ or from $K+1$th kernel by adding $\beta_KP_K(\zeta)P_K(\zeta')$.
The consistency of our result means the following identity.
\bea
&&\left(\frac{P^{(\nu)}_K(\zeta)P^{(\nu)}_{K-1}(\zeta')-P^{(\nu)}_K(\zeta')P^{(\nu)}_{K-1}(\zeta)}{\eta_{K}(\zeta-\zeta')}+ \frac{\alpha_K}{\eta_{K}} P_K(\zeta')P_{K}(\zeta) \right)
\\&&\quad=\left(\frac{P^{(\nu)}_{K+1}(\zeta)P^{(\nu)}_{K}(\zeta')-P^{(\nu)}_{K+1}(\zeta')P^{(\nu)}_{K}(\zeta)}{\eta_{K}(\zeta-\zeta')}+ \frac{\beta_{K}}{\eta_{K}} P_{K}(\zeta')P_{K}(\zeta) \right).
\eea
This is reduced to showing $\alpha_K-\beta_K=1$ which is easily obtained from the identity $u_K\ell_K=1$ and 
\be
\det[\A_K(0),\A_K'(0)]= \frac 1 {\det[\B_{K+1}'(0),\B_{K+1}(0)]},
\ee
which follows from a direct computation using the Schlesinger transformation.

There remains only one very exceptional case to consider: indeed both formul\ae\ in (\ref{FG}) may fail if the denominator appearing there vanishes.
This can happen since at $\varkappa = K\pm \frac 12$ one of the parameters $\ell,u$ does not tend to zero. 

However such unlucky situation does not occur with our construction in (\ref{explicit1}). 
A direct computation shows that
\be
\det[\B_{K+1}'(0),\B_{K+1}(0)] = -i\frac {(b-a)^{2K+1} (t_0^2-1)^{4K+2}}{ {t_0}^{6K+2r+6} 4^{2K+1}}.\label{detB}
 \ee
 Plugging into the denominator of (\ref{FG}) we have 
 \be
 1 + \ell_{_{K}}  \det[\B_{K+1}'(0),\B_{K+1}(0)] = 1+ 2i\pi \frac{C_0^{2\gamma K - \gamma}}{\eta_{_K}}    \det[\B_{K+1}'(0),\B_{K+1}(0)]
 \ee
Since $\eta_{K}$ is the norm square of the monic Freud polynomial and given the sign of the imaginary part of (\ref{detB}), the denominator is strictly positive.

\subsection{Arbitrarily improved error bound}
\label{arbitrary}

Here we explain how to construct the outer and the local parametrix that has arbitrary small error bound as one wants to achieve.  
Though we will not explicitly carry out the evaluation of the corrections,  the method already yields quite interesting identities which are otherwise hard to see.
The main idea is to generalize the Schlesinger transformation to a higher order. 
{\br
The general framework for arbitrary improvement of the error is not new and appeared in (\cite{DKMVZstrong}, Sec. 7.2), based on the inversion of an operator close to the identity in terms of a Neumann (geometric) series. The approach of \cite{Claeys} to the problem was indeed based on those ideas. In a certain sense our approach is a manipulation whose ``philosophical'' meaning is the same as computing the terms of the above--mentioned Neumann series, although the practical details may be different.
\er}

The {\it Weyl function} $W(\zeta)$ is defined by
\begin{equation}
W(\zeta):=\frac{1}{2\pi i}\int\frac{\d\mu(\xi)}{\xi-\zeta}=:-N^{-2\kappa\gamma}\sum_{j=1}^\infty\frac{\mu_j}{z^j}.
\end{equation}
(Let us write $z$ instead of the correct $\wt z$ as we are not going to deal with the physical coordinate.) The measure was given by $\d \mu(\xi)=\exp(-\xi^{2\nu+2})\d\xi$, but it can be general in the following discussion.
$\{\mu_j\}$ are the set of numbers defined by the expansion around $z=\infty$.  Note that they are also scaling with $N$ as $\mu_j\propto N^{2\kappa\gamma-\gamma j}$.

Then the matrix,
\be \left[\begin{array}{cc}1&N^{2\kappa\gamma}W(\zeta)
\\0&1\end{array}\right],\ee
satisfies the jump condition (\ref{Hjump}) for the local parametrix.
Since the jump property remains by any multiplication to the left, we may multiply a matrix to get:
\be\label{wtR} 
\wt R_R:=\left[\begin{array}{cc}1&\sum_{j=1}^R \mu_j/z^j
\\0&1\end{array}\right]
\left[\begin{array}{cc}1&N^{2\kappa\gamma}W(\zeta)
\\0&1\end{array}\right]={\bf 1}+{\cal O}(N^{2\kappa\gamma-\gamma(R+1)}),
\ \ z\in \pa\D
\ee
which can be made as close to the identity (on the boundary of the disk) as one wishes by increasing $R$.
Especially at $R=2K$ we get the most modest error bound $N^{-\gamma+2\gamma\delta}$, which we obtained in (\ref{errorestimate2}).

The same error bound can be obtained in a different way, using the following property of the Weyl function and the Pad\'e\ approximation provided by the orthogonal polynomials associated to the measure $\d\mu(\xi)$
\begin{equation}
W(\zeta)=-\frac{Q_k(\zeta)}{P_k(\zeta)}+{\cal O}(\zeta^{-2k-1}),
\end{equation}
where 
\be Q_k(\zeta):=\frac{1}{2i\pi}\int\frac{P_k(\zeta)-P_k(\xi)}{\zeta-\xi}\d\mu(\xi),
\ee is a polynomial of order $k-1$.

Then the following matrix also has the same error bound as $\wt R_{2K}$.
\be 
\hat R_K:=\left[\begin{array}{cc}1&N^{2\kappa\gamma}Q_K(\zeta)/P_K(\zeta)
\\0&1\end{array}\right]
\left[\begin{array}{cc}1&N^{2\kappa\gamma}W(\zeta)
\\0&1\end{array}\right]={\bf 1}+{\cal O}(N^{-\gamma+2\gamma\delta}).\ee

Now let us find the corresponding outer parametrix.
Here it is necessary that we start from $\Psi_0$ which does not have any pole at the outpost.
Defining $\Psi_0:=[\A_0,\B_0]$ we propose the outer parametrix of the form:
\be \label{wtpsi}
\wt\Psi_R:=\le({\bf 1}+\sum_{j=1}^R\frac{F_j}{z^j}\ri)[\A_0,\B_0].\ee
For $\wt\Psi_R \wt R_R$ to be analytic near $z=0$ we demand
\be\label{yeah}
\le({\bf 1}+\sum_{j=1}^R\frac{F_j}{z^j}\ri)[\A_0,\B_0]\left[\begin{array}{cc}1&\sum_{j=1}^R \mu_j/z^j
\\0&1\end{array}\right]={\cal O}(1),
\ee
which completely determines $\{F_j\}$.
As shown in the appendix, they are given by the solution of a linear equation.   

Now that we have explained the method to obtain an arbitrarily good error bound (on the boundary of the disk around the outpost) let us deduce a few implications.

As we have seen already, the leading outer parametrix is given by the $N$--independent $\Psi_K$; viceversa we have just defined a set of $N$--dependent outer parametrices $\wt\Psi_R$ for $R\geq 2K$ that should all converge to $\Psi_K$, i.e.
\be
\lim_{N\rightarrow\infty} \le({\bf 1}+\sum_{j=1}^R\frac{F_j}{z^j}\ri) \Psi_0=\Psi_K,\qquad R\geq 2K.
\ee
Remember that $\{F_j\}$ are determined by (\ref{yeah}).

Moreover, looking at (\ref{wtR}), the local parametrix does not contribute to the first column of the full asymptotic solution $\wt\Psi_R\wt R_R$ since $\wh R_K$ (or $\wt R_R$) are upper--triangular.  Therefore, all the information about the asymptotics of the orthogonal polynomials near the outpost in this setting is encoded directly in the outer parametrix itself (\ref{wtpsi}).

\section{(New) Universality behavior}
We can now examine the result we have obtained in the neighborhood of the outpost.

In particular we want to point out the behavior of zeroes of the  the orthogonal polynomials $p_n(z)$ in this particular scaling regime.
\begin{itemize}
\item The normalized counting measure of the  zeroes $\{z_1^{(n)} ,\dots, z_n^{(n)}\}$ of $p_n(z)$
\be
\nu_n(x):= \frac 1 n \sum_{k=1}^n \delta(x-z_1^{(n)})
\ee
 converges in the sense of measures to the usual equilibrium measure, namely for any continuous function $f(x)\in \mathcal C^0(\R)$
\be
\int_\R \nu_n(x) f(x) =  \frac 1 n \sum_{k=1}^n f(z_1^{(n)}) \to \int_{\R} f(x) \rho(x) \d x
\ee
\item The {\bf fine behavior} of those zeroes is precisely that in the scaling $s:= (C_0N)^{-\gamma} (x-\xi_0)$. For $0\leq\delta\leq 1/2$ they are converging to the location of the zeroes $\{s_1^{(K)}, \dots, s_K^{(K)}\}$ of the orthogonal polynomial of degree $K$ of the measure ${\rm e}^{-s^{2\nu+2}}\d s$. In measure--theoretical terms, for any compactly supported continuous  $f(s)$
\be
\int n \nu_n(x) f( (C_0 N)^\gamma (x-\xi_0)) =  \sum_{j=1}^K f(s_j^{(K)}).
\ee
\item  For $-1/2<\delta<0$ they are converging to the location of the zeroes $\{s_1^{(K-1)}, \dots, s_{K-1}^{(K-1)}\}$ of the orthogonal polynomial of degree $K-1$ of the measure ${\rm e}^{-s^{2\nu+2}}\d s$. (There is also the unique "stray zero" that scales as $x_{\mbox{stray}}\sim N^{-\gamma+2|\delta|\gamma}s_{\mbox{stray}}$.  However, the constant $s_{\mbox{stray}}$ is not universal.)
\item the correlation functions are the same  --in the same scaling- as for  the random matrix model of size $K$
\be
\d\mu(H_K):=
 {\rm e}^{- {\mathrm {Tr}}(H_K^{2\nu+2})}\d H_{K}
\ee
In a certain picturesque sense, there is a ``microscopical'' matrix model in the macroscopic background.
\item The {\bf kernel} for the correlation functions $K_{n}(x,x') = \frac {p_n(x)p_{n-1}(x')-p_{n-1}(x)p_n(x')}{h_{n-1}(x-x')}$ can be computed from 
\be
K_{n}(x,x') = \frac 1{2i\pi} \frac {[Y^{-1}(x') Y(x)]_{21}}{x-x'}
\ee
and thus a direct computation gives the new universal kernel near the outpost (in the scaling coordinate):
\be
K_n(\zeta',\zeta)=
-\frac{{\rm e}^{-\frac NT\le(g(z)+g(z')+\ell\ri)}}{N^{2\gamma\varkappa} (NC_0)^{\gamma}}
\left(\frac{P^{(\nu)}_K(\zeta)P^{(\nu)}_{K-1}(\zeta')-P^{(\nu)}_K(\zeta')P^{(\nu)}_{K-1}(\zeta)}{\eta_{K-1}(\zeta-\zeta')}\right)
\big(1+{\cal O}(N^{-\gamma+2\gamma|\delta|})\big),
\ee
At $\delta=\pm 1/2$ there appears a new term -- $P^{(\nu)}_{K}(\zeta')P^{(\nu)}_K(\zeta)$ or $P^{(\nu)}_{K-1}(\zeta')P^{(\nu)}_{K-1}(\zeta)$ depending on the sign -- appears.
\end{itemize}
As we see, rather surprisingly, a finite--size matrix model (of particular type) arises naturally as a scaling limit of a general one.

To conclude we remark that using the same methods employed here it is possible to handle a similar multicritical phenomenon where the ``microscopic'' matrix model has {\em any} polynomial exponential weight rather than the Freud one. This requires, however, a more finely--tuned potential, namely we need to introduce a dependence on $N$ into $V$ and $T$. Similar considerations, but leading to Freud weights as in the main body of the text, are contained in Appendix \ref{App1}.

We postpone this analysis to a future publication.
\appendix
\renewcommand{\theequation}{\Alph{section}.\arabic{equation}}
\section{Double--scaling approach}
\label{App1}

In this section we explain why our simplified approach using the chemical potential is {\em de facto} equivalent to a more refined double--scaling limit.
{In \cite{MoBirth, Claeys} the deformation of the $g$--function was written in terms of the parameter $T$ by fine-tuning $T-T_{cr}$ ($T_{cr}$ being the critical total mass of the measure): in \cite{MoBirth} a small dependence on $V$ was also introduced so that the critical point in the effective potential remains of the same order of vanishing,  $2\nu+2$. Here we want to illustrate how our simplified approach can be related to those.}

Suppose that $V:= V_\epsilon(x),T:=  T_\epsilon$ depend on an external parameter which we denote by $\epsilon$ in such a way that
there exist a (finite) disjoint union of  bounded intervals $J = \sqcup J_k$ with endpoints smoothly depending on $\epsilon$, a point $\xi_0(\epsilon)\not \in J$ (also smoothly depending on $\epsilon$   and a real function $h(x)$ with the properties

\begin{wrapfigure}{r}{0.5\textwidth}
\resizebox{0.4\textwidth}{!}{
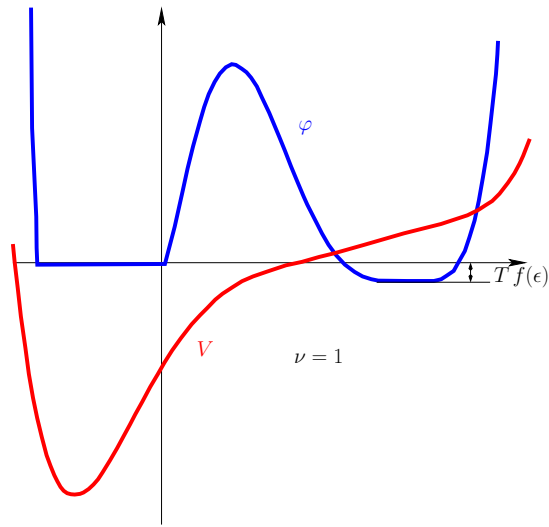}
\caption{An example with $\nu=1$ and $V$ of degree $6$. This is a numerically correct plot, although the axis are scaled differently for $V$ and $\varphi$.}
\label{CritPot}
\end{wrapfigure}

\begin{itemize}
\item $h(x)$ is harmonic $\C\setminus J$ and continuous in $\C$ and  at infinity $h(x) \sim \ln |x|$;
\item $\frac 1 T V (x) - h(x) \equiv 0$ for $x\in J$;
\item $\frac 1 T V(x)-h(x) = C(\epsilon) (x-\xi_0)^{2\nu+2} (1+ \mathcal O(x-\xi_0))- f(\epsilon)$, with $f(\epsilon)$ a smooth function in  a neighborhood of $\epsilon=0$ with  $\R\ni f'(0)\neq 0$;
\item other than the negative sign implied by the previous bullet-point in a neighborhood of $\xi_0$ (for small $\epsilon$) the sign of $V(x)-h(x)$ on $\R\setminus J$ is strictly positive (see Fig. \ref{CritPot}).
\item  The sign of $\frac 1 T\Re V(z) - h(z)$ is negative on a left/right neighborhood of each component of $J$, and the size of this neighborhood is uniform in $\epsilon$ for small $\epsilon$'s.
\end{itemize}
In the above $C>0$ may depend on $\epsilon$ as long as  it is smooth and bounded away from $0$, and all the Landau symbols should be uniform in $\epsilon$. The function $h$ here is nothing but the real part of the $g$--function (up to addition of the Robin constant) and the last bullet-point is equivalent to saying that $h$ is the logarithmic energy of a {\em positive} measure supported on $J$ (a consequence of Cauchy--Riemann equation for harmonic functions).
In this framework then we could repeat verbatim the analysis by fine--tuning $\epsilon$ via the implicit equation 
\be
f(\epsilon) =  {2\varkappa\gamma}\frac {\ln N}{N}
\ee
This equation defines $\epsilon(N)$ (for suitably large $N$) since $f(\epsilon)$ has nonzero derivative at $\epsilon=0$ and hence invertible near $0$.

All the construction would still apply verbatim with the caveats that 
the turning points are (slowly) moving in $N$ and hence the outer parametrix, the local coordinates $\zeta$ used to define the local parametrices (including the one at the outpost $\xi_0$) depend (smoothly and slowly) on $N$\footnote{By ``smoothly and slowly'' we mean that they are smooth functions of $\epsilon$ which --in turn-- is of order $\ln N/N$}. 

The situation is not dissimilar to the one in \cite{KenArnoLaguerre} where an $N$--dependent $g$--function (with $N$--dependent turning points) was employed. None of the analysis we have carried out is significantly affected. In particular the local RHP for the parametrix at the outpost is identical with the only understanding that the conformal parameter $\zeta(z)$ depends on $N$ but only through $\epsilon$ and hence in a uniformly bounded way.

To convince the reader that the above list of requirements is not insanely restrictive we show how to construct one such family of potentials $V$ and total charges $T$. For simplicity we restrict to a one--cut situation but this is purely in the interest of conciseness. Methods similar to \cite{BertoBoutroux} could be used to construct a family for an arbitrary number of cuts.

Suppose that $V(x)$ is a (real) {\em polynomial} of even degree and that $V_{0},T_{0}$ is a multicritical pair as the one used in the main text. It is not difficult to show that we must have in general $\deg V \geq 2\nu+4$.

We now define a {\bf deformation} depending on the parameter $\epsilon = T-T_{cr}$. Here we have chosen for transparency of exposition the deformation parameter as the deviation from the critical total charge, but in  general it may be an abstractly introduced parameter.

As explained for example in \cite{BertoBoutroux} we can write the (complex) effective potential $\varphi_{0}$ as
\be
\varphi_{0}(z)  = \int_{b_0}^z M_0(x)(x-\xi_0)^{2\nu+1} \sqrt{(x-a_0)(x-b_0)},\ \ a_0<b_0<\xi_0.
\ee
where $\deg M_0 = \deg V' - 2\nu-2$ and it is determined by the (algebraic) equation 
\be
M_0(z) (z-\xi_0)^{2\nu+1} \sqrt{(z-a_0)(z-b_0)} = V_0'(z) - \frac  {T_0} z + \mathcal O(1/z^2)\ .
\ee
For simplicity in what follows we assume that the roots $\mu_{j,0}$ of $M_0(z)$ are simple.
It is clear that $V_0$ cannot be a totally arbitrary potential (a simple parameter counting confirms this).

The criticality condition amounts to 
\be
 \int_{b_0}^{\xi_0} M_0(x)(x-\xi_0)^{2\nu+1} \sqrt{(x-a_0)(x-b_0)} = 0
\ee
which implies that $M_0(x)$ has an odd number of zeroes in $(b,\xi_0)$\footnote{Since $\deg M_0\geq 1$ the assertion on the minimal degree of $V$ follows.}. Other than this (unless other critical phenomena occur) $M_0(x)>0 $ on $[a,b]$ and $M_0(\xi_0)>0$.

Define now 
\be
P_0 (z):=  M_0(x)^2(x-\xi_0)^{4\nu+2}{(x-a_0)(x-b_0)}:
\ee
we will define a first--order ODE for $P_\epsilon$ with Cauchy data $P_\epsilon|_{\epsilon=0} = P_0$. 
Since we want to preserve the multiplicity of the root at the outpost $\xi_\epsilon$  we must have $\dot P_\epsilon = (z-\xi_\epsilon)^{4\nu+1}(1+...)$ and hence $\dot P_\epsilon$ must be {\em at least} a polynomial of that order; moreover, since it should also preserve the square of $M_\epsilon$, it must be divisible by $M_\epsilon$.  The simplest choice is to take directly 
\be
\dot P_\epsilon  := A^{-1} M_\epsilon(z) (z-\xi)^{4\nu+1}\ .\label{ODE}
\ee
The constant $A$ is determined by the requirement that $\res{\infty} \frac{\dot P}{2\sqrt{P}} =  \dot T= 1$, namely 
\be
 \res{z=\infty}  \frac {(z-\xi)^{2\nu}}{2\sqrt{(z-a)(z-b)}} =  A
\ee

This determines $A$ as a (rather cumbersome)  {\bf polynomial} in $\xi,a,b$: note that since we have chosen the determination of the squareroot that is {\em positive} on the real positive axis near $\infty$, then $A<0$.

The potential $V_\epsilon$ undergoes the evolution according to 
\be
\frac{\dot P_\epsilon}{2\sqrt{P_\epsilon}} = \dot V'_\epsilon(z) - \frac 1 z + \mathcal O(z^{-2}).
\ee
Hence the coefficients of $V_\epsilon$ up to degree $2\nu$ will necessarily depend on $\epsilon$. Only for the simplest case of $\nu=0$ we can keep the potential fixed.

The full ODE  is thus either eq. (\ref{ODE}) or --in terms of the position of the zeroes of $P$--
\bea
\dot \xi_\epsilon &\&=    -\frac {\dot P_\epsilon ^{(4\nu+1)}(\xi)}{(4\nu+2)P_\epsilon^{(4\nu+2)}(\xi)} = \frac {-A^{-1}}{(4\nu+2)M(\xi)(\xi-a)(\xi-b)} \ ,\cr
\dot a  &\&= -\frac {\dot P_\epsilon(a)}{P'(a)} = \frac {-A^{-1}} {M(a) (\xi-a)(a-b)} \ ,\qquad
\dot b  = -\frac {\dot P_\epsilon(b)}{P'(b)} = \frac {-A^{-1}}{M(b) (\xi-b)(b-a)}\cr
\dot \mu_{j} &\&= -
\frac{\dot P_\epsilon'(\mu_{j})}  {2P''(\mu_j)} 
\eea

The theorem of existence for ODE guarantees that the above equation has solution for $\epsilon$ in a suitable interval around $\epsilon=0$ (with  $P_0$ as IVP).

The only point we verify in addition is the claim about the behavior of the effective potential near the outpost $\xi$; but this is an elementary application of Taylor theorem since (noting that $\varphi_\epsilon(b_\epsilon)\equiv 0$)
\be
\varphi_\epsilon(x) = \overbrace{\int_b^\xi \varphi'_\epsilon(s)\d s}^{=:  T f(\epsilon)} +\frac{ M_\epsilon(\xi)\sqrt{(\xi-a)(\xi-b)}}{2\nu+2} (x-\xi)^{2\nu+2}(1 + \mathcal O(x-\xi))
\ee
The derivative of $f(\epsilon)$ at $\epsilon=0$ is 
\be
\dot f(0)= \frac 1{T_0}\int_{b_0}^{\xi_0} \frac  {A(x-\xi_0)^{2\nu}}{\sqrt{(x-a_0)(x-b_0)}} \d x <0
\ee

This implies that for $\epsilon>0$ ($T>T_ {cr}=T_0$) the effective potential is {\em negative} in a small neighborhood of the outpost $\xi$, with a minimum value that should be fine--tuned as detailed above.

\section{Construction of the outer parametrix for arbitrary number of cuts}
\label{multicut}

We only sketch the construction since the details would require a good deal of notation to be set up. We will use the same notation and ideas contained in \cite{BertolaMo}.

We denote by $w$ double-cover of the $z$--plane branched at the endpoints of the support of the equilibrium measure
\be
w^2:= \prod_{j=1}^{2g+2} (z-\alpha_j)
\ee
This is a hyperelliptic algebraic curve of genus $g$. We denote by $\infty_{\pm}$ the two points above $z=\infty$ in the usual compactification of the curve, and by $p_{\pm}$ the two points projecting to the location of the outpost\footnote{The point $\infty_+$ is characterized by $w>0$ as $z\in \R_+$ near $\infty$. The point $p_+$ is the point on the Riemann surface of $w$ obtained by analytic continuation of $w$ on the complex plane slit along $(\alpha_{2j-1},\a_{2j})$}.
We denote by $\omega_j$ the first--kind differentials normalized along the $a$--cycles: explicitly 
\be
\omega_j(z) = \sigma_{j\ell} \frac {z^{\ell-1} \d z}{w}\ ,
\ee
where the summation over repeated indices is understood (and they range from $1$ to $g$) and $\sigma_{j\ell}$ is an invertible matrix such that $\oint_{a_k} \omega_j = \delta_{jk}$. 

Using the standard notation for divisors on Riemann surfaces \cite{FarkasKra} we consider the unique (up to multiplicative constant) sequence of spinors with the divisor properties
\bea
&& (\psi_r^{(0)})\geq -(r-1)\infty_+ + r\infty_- + K(p_+-p_-)\ ,\qquad r\in \Z\\
&& \psi_r^{(0)\star}(p) := \psi_r^{(0)}(p^\star) 
\eea
where $p\mapsto p^\star$ is the holomorphic involution of the hyperelliptic curve.
The spinors (and their starred counterparts) are also sections of the line bundles $\L, \L^{-1}$ with character $\chi$ ($\chi^{-1}$ respectively) defined by
\bea
\chi(\gamma):= \le\{
\begin{array}{cc}
\ds {\rm e}^{2i\pi \mathcal A_j}:= \prod_{\ell=1}^{j} {\rm e}^{iN\epsilon_{\ell}} & \hbox{ for } \gamma = a_j\\
{\rm e}^{2i\pi \mathcal B_j}:=1 & \hbox{ for } \gamma = b_j.
\end{array}
\ri.\cr
\epsilon_j:=\frac 2 T \int_{\alpha_{2j-1}}^{\alpha_{2j}} \rho(x)\d x = \hbox { $j$-th {\bf filling fraction}}
\eea

These are the generalization to arbitrary genus of the spinorial Baker--Akhiezer vector used earlier:  the {\bf characteristics} $\vec{\mathcal A}, \vec{\mathcal B} \in \C^g$ are defined up to integers.

The matrix
\be
\Psi_{r,K} := \frac 1{\sqrt{\d z}} \le[
\begin{array}{cc}
\psi_{r,K}(p)  &i \psi_{r,K}(p^\star)\\ 
-i \wt \psi_{r-1,K}(p)  & \wt \psi_{r-1,K}(p^\star)\\ 
\end{array}
\ri]\bigg|_{p=p^{-1}(z)}
\ee
solves the model RHP with quasi--permutation monodromies on the cuts and on the gaps (antidiagonal on the cuts, diagonal on the cuts). Here $p^{-1}(z)$ is the point $(z,w_+(z))$  where $w_+(z)$ is determination of $w(z)$ that behaves like $z^{g+1}$ at infinity, analytically extended to the complex plane sliced along the support of the equilibrium measure (i.e. the {\bf physical sheet}).

The spinor $\sqrt{\d z}$ is defined on the {\em double cover} of the hyperelliptic curve (it has $\sqrt{\phantom{-}}$ branch-behavior at the Weierstrass points in terms of the local parameter $\sqrt{z-\alpha_j}$, hence has singularities of type $\sqrt[4]{z-\alpha}$ when thought of as a spinor on the plane).

\begin{wrapfigure}{r}{0.6\textwidth}
\resizebox{0.6\textwidth}{!}{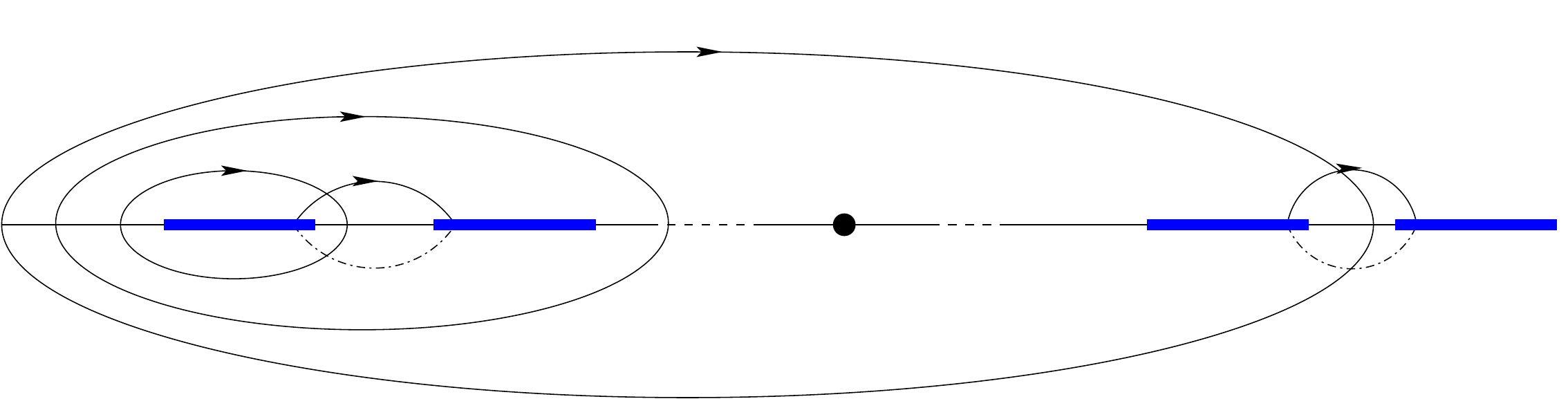}
\end{wrapfigure}
The first column has a {\bf zero} of order $K$ at $z(p_+)=\xi_0$ (the outpost) and the second column a pole of order (at most) $K$: at infinity it behaves as $z^{r\sigma_3}$ (up to left multiplicative constants).

 The expression in terms of $\Theta$ functions is 
\bea
\varphi_{r,K}&\& :=  \frac {\Theta_{_\Delta}^r(p-\infty_-)\Theta_{_\Delta}^K(p-p_+) \Theta\le[\mathcal A \atop 0 \ri] \le(p+r\infty_- - (r+1)\infty_+ + K(p_+-p_-)\ri)}
{ \Theta_{_\Delta}^{r+1}(p-\infty_+)\Theta_{_\Delta}^K(p-p_-)} \frac{h_\Delta(p)}{\sqrt{\d z}}\\
\psi_r &\&=\frac {\varphi_{r,K}(p)}{ C_{r,K}}\ ,\qquad\wt \psi_r  =  \frac {\varphi_{r,K}(p)}{\wt C_{r,K}} \\
&\& C_{r,K} = \lim_{p\to \infty_+} \frac {\varphi_{r,K}}{z^{r}\sqrt{\d z}}\ ,\qquad  \wt{C_{r,K}} = \lim_{p\to \infty_-} \frac {\varphi_{r,K}}{z^{r}\sqrt{\d z}} \\
&\& h_\Delta := \sqrt{\sum_{j=1}^g \pa_{z_j} \Theta_{_\Delta}(0) \omega_j}
\eea
The notation (rather standard) is lifted from \cite{BertolaMo} and \cite{Faybook}: the Abel map is understood when writing points as arguments of $\Theta$ and it is based at one of the Weierstrass points (for example $\a_1$) $\Delta$ is an arbitrary odd non-singular half-period.
Recall that (pag. 23 of \cite{Faybook}) all such  characteristics $\Delta$ are in one-to-one correspondence with partitions of the Weierstrass points into $g-1$ and $g+3$ points $\{\a_{k_1},\dots, \a_{k_{g-1}} \}\sqcup\{\a_{\ov k_1} ,\dots, \a_{\ov k_{g+3}} \}$
It is to be noted that (pag. 23 of \cite{Faybook})
\be
\frac{h_\Delta(p)}{\sqrt{\d z}} =
\sqrt{-\pa_\ell \Theta_{_\Delta}(0) \sigma_{\ell g}}
\frac {\sqrt[4]{ \prod_{\ell=1}^{g-1}(z-\a_{k_\ell})} }{
 \sqrt[4]{\prod_{\ell=1}^{g+3}(z-\a_{\ov k_\ell})}}
\ee
and that 
\be
\frac{ \Theta\le[\mathcal A \atop 0 \ri] \le(p+r\infty_- - (r+1)\infty_+ + K(p_+-p_-)\ri)h_\Delta(p)}{\Theta_{_\Delta}(p-\infty_+)\sqrt{\d z}}  = \frac {F(z)}{\sqrt[4]{\prod_{j=1}^{2g+2} (z-\a_j)}} 
\ee
where $F(z)$ is an analytic function with jump discontinuities on the cuts\footnote{
Of course this is a bit vague description since we should first stipulate how the fourth-roots have been defined.
} and it is {\em independent} of the choice of $\Delta$, it is bounded in the finite region of the $z$--plane and growing like $z^{\frac{g+1}2}$ at infinity.
Moreover, straightforward computations show that  (with some {\em overall } ambiguity of signs)
\bea
 \Theta_{_\Delta}(p-\infty_\pm ) \mathop{\sim}_{p\to \infty_{\pm}} \mp \frac 1 z\pa_\ell \Theta_{_\Delta}(0) \sigma_{\ell g}\\
  \frac{h_\Delta(p)}{ \Theta_{_\Delta}(p-\infty_\pm)\sqrt{\d z}}\mathop{ \longrightarrow }_{p \to \infty_{\pm}} \le(\mp \pa_\ell \Theta_{_\Delta}(0) \sigma_{\ell g} \ri)^{-\frac 12} \\
 C_{K,r} = \frac {
 \Theta_{_\Delta}^r(\infty_+-\infty_-)  \Theta_{\Delta}^{K}(\infty_+-p_+) \Theta\le[\mathcal A \atop 0\ri] \le(r(\infty_- - \infty_+) + K(p_+-p_-)\ri) }{
 (-\pa_\ell \Theta_{_\Delta} \sigma_{\ell g})^{r+\frac 12 } \Theta_{\Delta}^{K}(\infty_+-p_-) }\\
 \wt{C_{K,r}}  = \frac {(\pa_\ell \Theta_{_\Delta} \sigma_{\ell g})^{r+\frac 12 }
  \Theta_{\Delta}^{K}(\infty_--p_+)  \Theta\le[\mathcal A \atop 0 \ri] \le((r+1)(\infty_- - \infty_+) + K(p_+-p_-)\ri)}{ \Theta_{_\Delta}^{r+1}(\infty_--\infty_+) \Theta_{\Delta}^{K}(\infty_--p_-) }
\eea

The constant $C_{r,K}$ is simply the normalization so that $\psi_{r}$ behaves as $z^r$ at $\infty_+$ while $\wt C_{r,K}$ is chosen so that $\wt \psi_{r}(p^\star)$ behaves as $z^{-r-1}$ at $\infty_+$.
\br
The fact that these formal expression do not vanish identically follows from the fact that 
\be
\Theta\le[\mathcal A \atop 0 \ri] (r(\infty_- - \infty_+) + K(p_+-p_-)) \neq 0
\ee
for arbitrary $\mathcal A\in \R^g$, as these correspond to positive divisors of degree $g$ with $g$ points in the gaps. This is proved in a more general setting in [Prop. 6.3, pag. 111 of \cite{Faybook}]. Of course for $K=0=r$ the nonvanishing of this expression is precisely the same that appears in \cite{DKMVZ} although maybe not clearly stated. 

From a point of view of isomonodromic theory, the above theta function is intimately related to the isomonodromic tau function \cite{JMU}, whose vanishing determines the (non)solvability of a Riemann--Hilbert problem. In this example the RHP is the model problem for the OPs.  If the spectral curve had no real-structure then in general it could happen that for exceptional values the problem does not admit a solution (see \cite{BertolaMo})
\er

We conclude this section with a few important remarks and shortcomings of these formulas
\begin{itemize}
\item the construction of the improved parametrix in Sect. \ref{sectimproved} did not use the specific form of the outer parametrix but just the jet-expansion near the outpost, thus it applies {\em verbatim} to the general case, with the proviso of the next point;
\item the description of the behavior of the roots at the outpost remains generically valid in this case: however the {\em direction} of approach of the stray zero (in eq. (\ref{localstray})) depends on the actual sign of the expressions involved, hence in this general case it cannot be easily identified\footnote{
Nor it should be expected to always come from one side. On a heuristic level, the stray zero should come from the ``closest'' spectral band, and hence it depends on the location of the outpost.};
\item for the Stokes' values $\varkappa \in \Z+\frac 12$ there is the potential for the denominators of formul\ae\ (\ref{FG}) to vanish under exceptional circumstances (i.e. for special spectral curves and special values of $K$). This would make the approximation (\ref{firstcol}) unbounded in $N$ and hence invalidate it. For the one--cut case as in the main text it was rather simple to directly  verify  that the determinants in (\ref{FG}) have a suitable sign so that the denominators are bounded away from zero, but for the case of multi-cut solutions a similar computation requires a deep manipulation of $\Theta$ functions and we could not determine a similar property. We suspect that such property should hold here too on account of the reality conditions of the cuts and the Jacobian of the spectral curve.
\end{itemize}

\section{Asymptotics: long results}
\label{fullasymptotics}
We write the full asymptotics for the first column of $\wt Y$ up to  ${\cal O}(N^{-2\gamma})$.
\bea 
&&\left(N^{\gamma}C_0^{\gamma}\right)^K \Psi_\varkappa R_\varkappa\big|_{(1)}
\\&&\nonumber
=P^{(\nu)}_K(\zeta)\left(\wt\A_K+\frac{u_K\det[\wt\A_K,\wt\A_K(0)]}{1+u_K\det[\wt\A_K(0),\wt\A_K'(0)]}\frac{\wt\A_K(0)}z-\frac{\ell_{K-1}\det[\wt\A_K,\wt\B_K(0)]}{1+\ell_{K-1}\det[\wt\B_K'(0),\wt\B_K(0)]}\frac{\wt\B_K(0)}z+\frac{\ell_{K-1}}z\wt\B_K\right)
\\&&\nonumber
\quad+P^{(\nu)}_K(\zeta)\le(\frac{\ell_{K-1}}{\wt z}-\frac{\ell_{K-1}}{z}\ri)
\le(\wt\B_K-\frac{\ell_{K-1}\det[\wt\B_K,\wt\B_K(0)]}{1+\ell_{K-1}\det[\wt\B_K'(0),\wt\B_K(0)]}\frac{\wt\B_K(0)}z\ri)
\\&&\nonumber
-\ell_{K-1}N^{\gamma}C_0^{\gamma}P^{(\nu)}_{K-1}(\zeta)
\left(\wt\B_K+\frac{u_K\det[\wt\B_K,\wt\B_K(0)]}{1+u_K\det[\wt\A_K(0),\wt\A_K'(0)]}\frac{\wt\A_K(0)}z-\frac{\ell_{K-1}\det[\wt\B_K,\wt\B_K(0)]}{1+\ell_{K-1}\det[\wt\B_K'(0),\wt\B_K(0)]}\frac{\wt\B_K(0)}z+\frac{u_{K}}z\wt\A_K\right)
\\&&\nonumber\qquad+{\cal O}(N^{-2\gamma}).
\eea
At half integer $\varkappa$ we have the following leading and the subleading behavior.
\bea&&
\left(N^{\gamma}C_0^{\gamma}\right)^K \Psi_\varkappa R_\varkappa\big|_{(1)}\simeq
\frac{\A_K(0)}{1+u_K\det[\A_K(0),\A_K'(0)]}P_K^{(\nu)}(\zeta)
\\\nonumber&&\qquad\qquad\qquad -\ell_{K-1}C_0^\gamma N^\gamma\left(\B_K(0)+\frac{u_K\det[\B'_K(0),\B_K(0)]}{1+u_K\det[\A_K(0),\A_K'(0)]}\A_K(0)+u_K\A_K'(0)\ri)P_{K-1}^{(\nu)}(\zeta)
\\\nonumber&&\qquad\qquad\qquad +\le(\A_K'(0)+\frac{u_K\det[\frac12\A''_K(0),\A_K(0)]}{1+u_K\det[\A_K(0),\A_K'(0)]}\A_K(0)\ri)\frac{\zeta P_K^{(\nu)}(\zeta)}{C_0^\gamma N^\gamma}+{\cal O}(N^{-\gamma}),
\\\nonumber&&
\left(N^{\gamma}C_0^{\gamma}\right)^K \Psi_\varkappa R_\varkappa\big|_{(1)}\simeq-\frac{\ell_{K-1}C_0^\gamma N^\gamma\B_K(0)}{1+\ell_{K-1}\det[\B_K'(0),\B_K(0)]}P_{K-1}^{(\nu)}(\zeta)
\\\nonumber&&\qquad\qquad\qquad
\le(\A_K(0)+\ell_{K-1}\B_K'(0)-\ell_{K-1}\frac{\det[\A_K'(0),\B_K(0)]-[z/\wt z]_1}{1+\ell_{K-1}\det[\B_K'(0),\B_K(0)]} \B_K(0)\ri)P^{(\nu)}_K(\zeta)
\\\nonumber&&\qquad\qquad\qquad
-\ell_{K-1}\le(\B_K'(0)-\frac{\ell_{K-1}\det[\B_K''(0),\B_K(0)]}{1+\ell_{K-1}\det[\B_K'(0),\B_K(0)]}\B_K(0)\ri)P_{K-1}^{(\nu)}(\zeta)\zeta+{\cal O}(N^{-\gamma}).
\eea
for $\varkappa=K+1/2$ and $\varkappa=K-1/2$, respectively.

\bibliographystyle{unsrt}
\bibliography{Colonization}
\end{document}